\documentclass{aa}  

\newdimen\imgwidth
\usepackage{graphicx}
\usepackage{txfonts}
\usepackage[usenames]{color}
\usepackage{longtable}
\usepackage{dcolumn}

\begin{document}
   \title{Line profile and continuum variability in the 
          very broad-line Seyfert galaxy Mrk\,926
\thanks{Based on observations obtained with the Hobby-Eberly
               Telescope, which is a joint project of the University
               of Texas at Austin, the Pennsylvania State
           University, Stanford University, Ludwig-Maximilians-Universit\"at
            M\"unchen, and Georg-August-Universit\"at G\"ottingen.}
 }

   \author{W. Kollatschny
          ,  M. Zetzl
          }

   \institute{Institut f\"ur Astrophysik, Universit\"at G\"ottingen,
              Friedrich-Hund Platz 1, D-37077 G\"ottingen, Germany\\
              \email{wkollat@astro.physik.uni-goettingen.de}
}


   \date{Received 13 October 2009; accepted 10 June 2010}
   \authorrunning{Kollatschny et al.}
   \titlerunning{Short-term variations in Mrk\,926}


 \abstract{}
{We present results of an intensive spectroscopic variability campaign of
the very broad-line Seyfert 1 galaxy Mrk\,926.
 Our aim is to investigate the broad-line region (BLR) by studying
 the intensity and line profile variations of this
 galaxy on short timescales.}
{High signal-to-noise ratio (S/N)
 spectra were taken with the 9.2m Hobby-Eberly Telescope (HET)
in identical conditions during two observing campaigns in 2004 and 2005.
After the spectral reduction and internal calibration we achieved
a relative flux accuracy of better than 1\%.}
{The rms profiles of the very broad Balmer lines 
have shapes that differ from their mean line profiles, consisting of
two inner (v $\lesssim \pm{}$ 6~000 km s$^{-1}$)
and two outer (v $\gtrsim \pm{}$ 6~000 km s$^{-1}$)
 line components in addition to a central component
(v $\lesssim \pm{}$ 600 km s$^{-1}$).
These outer and inner line segments 
varied with different amplitudes
during our campaign.
The radius of the BLR is very small with an
 upper limit of 2~light-days for the H$\beta$
 BLR size.
We derived an 
 upper limit to the  central black hole mass of
$ M= 11.2 \times 10^{7} M_{\odot} $.
The 2-D cross-correlation functions CCF($\tau$,$v$) of H$\beta$ and H$\alpha$
 are flat within the error limits. The response of the Balmer line
segments with respect to continuum variations is different in the
outer and inner wings of H$\alpha$ and H$\beta$.
This double structure in the 
response curves - of two separate inner and outer components -
has also been seen in the rms line profiles. We conclude 
that the outer and inner line segments
originate in different regions and/or under different physical conditions.
}
{}


   \maketitle
%

\section{Introduction}

We selected Mrk\,926 (MCG-2-58-22) as a high-priority target for a detailed
spectroscopic variability study with the 9.2m Hobby-Eberly Telescope (HET).
Mrk\,926 is a Seyfert 1 galaxy with very broad emission lines,
its Balmer and helium lines having widths (FWHM) of more than 10\,000
km s$^{-1}$. In their study of the long-term variability of very broad-line
AGNs, Kollatschny et al. (\cite{kollatschny06}) demonstrated Mrk\,926 has 
strong H$\beta$ line variability amplitudes.

Many details about the size, structure, and kinematics of the innermost
line-emitting
region of active galactic nuclei (AGNs) - the broad-line region - remain
unclear. In particular almost no very broad-line AGN has been studied so far
in detail.

The nuclear activity was first detected in 
Mrk\,926 by
Ward et al. (\cite{ward78}) when they acquired a spectrum of the optical
counterpart to the strong X-ray source.
Mrk\,926 is a compact source in optical images
(Garnier et al. \cite{garnier96}).
It is the brighter galaxy member
(RA=23:04:43.5, Dec=-08:41:09 (2000), z=0.04701)
in a double system. The close companion 2MASX~J23044397-0842114
(RA=23:04:44.0, Dec=-08:42:11 (2000), z=0.04731) is located one arc min
to the south.

Osterbrock \& Shuder (\cite{osterbrock82}) published emission-line profiles
of  Mrk\,926, reporting that the lines did not vary between July 1978  
and December 1979. Kollatschny et al. (\cite{kollatschny06}) demonstrated
that both continuum and Balmer lines
of Mrk\,926 varied strongly
over a period 
of eight years between 1990 and 1997. 

In this paper, we concentrate on the line profile variations
of Mrk\,926.
The line profiles of AGN and their variations can provide us information about
the structure of the line-emitting region.
We study in detail the variations in the individual line segments of
the Balmer lines and the changes in the profiles during a variability
campaign in 2005 and another shorter campaign performed one year
before in 2004. 

There are not that many Seyfert galaxies for which line profile variations
have been studied in greater detail than e.g., 
NGC5548 (Peterson et al., \cite{peterson02} and references therein, Wanders \&
  Peterson, \cite{wanders97}, Sergeev et al., \cite{sergeev07}, Denney et al.,
 \cite{denney09}, Bentz et al., \cite{bentz09}), 
NGC1097 (Storchi-Bergmann, \cite{storchibergmann03}), 
NGC3227 and NGC 3516 (Denney et al., \cite{denney09}),
NGC4151 (Penston \& Perez, \cite{penston84}, Sergeev et al., \cite{sergeev01}),
NGC4593 (Kollatschny \& Dietrich, \cite{kollatschny97}),
NGC4748 (Bentz et al., \cite{bentz09}),
NGC7603 (Kollatschny et al., \cite{kollatschny00}),
OQ208=Mrk668 (Marziani et al., \cite{marziani93}, Gezari et
al., \cite{gezari07}), 
Akn120 (Kollatschny et al., \cite{kollatschny81},  Peterson et al.,
 \cite{peterson98a}, Doroshenko et al., \cite{doroshenko08}),
F9 (Kollatschny \& Fricke, \cite{kollatschny85}), 
Arp102B and 3C332  (Gezari et al., \cite{gezari07}),
Arp151 (Bentz et al., \cite{bentz08}, Bentz et al., \cite{bentz09}),
3C390.3 (Sergeev et al., \cite{sergeev02}, Gezari et al., \cite{gezari07}),
SBS1116 (Bentz et al., \cite{bentz09}),
Mrk\,1310 (Bentz et al., \cite{bentz09}), 
and Mrk\,110 (Kollatschny et al., \cite{kollatschny01}).
Mrk\,110 is
the only galaxy for which a detailed two-dimensional
(2D) reverberation mapping study has been carried out so far
(Kollatschny \cite{kollatschny02}, \cite{kollatschny03}).

\section{Observations and data reduction}

 We acquired optical spectra of Mrk\,926 with the 9.2-m Hobby-Eberly
 Telescope (HET) at McDonald Observatory during 4 epochs between
 August 7 and October 9, 2004
 and 15 epochs between August 3 and December 6, 2005.
 The log of observations is given in Table 1.
\begin{table}
\tabcolsep+8mm
\caption{Log of observations}
\centering
\begin{tabular}{ccc}
\hline 
\noalign{\smallskip}
Julian Date & UT Date & Exp. time \\
2\,400\,000+&         &  [sec.]   \\
\hline 
53224.362 & 2004-08-07 & 900.0 \\
53238.332 & 2004-08-21 & 900.0 \\
53258.283 & 2004-09-10 & 900.0 \\
53287.194 & 2004-10-09 & 900.0 \\
53585.380 & 2005-08-03 & 900.0 \\
53590.362 & 2005-08-08 & 900.0 \\
53600.329 & 2005-08-18 & 900.0 \\
53609.318 & 2005-08-27 & 900.0 \\
53628.262 & 2005-09-15 & 900.0 \\
53629.255 & 2005-09-16 & 900.0 \\
53644.212 & 2005-10-01 & 900.0 \\
53653.194 & 2005-10-10 & 900.0 \\
53670.147 & 2005-10-27 & 900.0 \\
53674.144 & 2005-10-31 & 900.0 \\
53680.109 & 2005-11-06 & 900.0 \\
53686.096 & 2005-11-12 & 900.0 \\
53693.090 & 2005-11-19 & 900.0 \\
53700.070 & 2005-11-26 & 900.0 \\
53710.040 & 2005-12-06 & 900.0 \\
\hline 
\end{tabular}
\end{table}
We obtained 19 spectra over a period of 485.7 days.
Fifteen of these spectra were taken over a period of 124.7 days in 2005.
The average interval
between these observations was 8.9 days and the median interval was 7.9 days.
In some cases we acquired spectra at intervals of only a few days.

All observations were performed in identical instrumental conditions with the
Marcario Low Resolution Spectrograph (LRS)
mounted at the prime focus of HET. The detector was
a 3072x1024 15 $\mu$m pixel Ford Aerospace CCD with 2x2 binning. 
The spectra cover the wavelength range from 4200\,\AA\
to 6900~\AA\ (LRS grism 2 configuration)
 in the rest frame of the galaxy
with a resolving power of 650 at 5000\,\AA\ (7.7\,\AA\ FWHM).
All observations were taken with an exposure time of 
15 minutes, which in most cases yielded a     
S/N  of at least 100
per pixel in the continuum.
The slit width was fixed to
2\arcsec\hspace*{-1ex}.\hspace*{0.3ex}0 projected on the
sky at an optimized position angle to minimize differential refraction.
Furthermore, all observations were taken at the same airmass
thanks to the particular design feature of the HET.
We extracted 7 columns from each of our object spectra 
corresponding to 3\arcsec\hspace*{-1ex}.\hspace*{0.3ex}3. The spatial
resolution was 0\arcsec\hspace*{-1ex}.\hspace*{0.3ex}472 per binned pixel.

Both HgCdZn and Ne spectra were taken after each object exposure
to enable a
wavelength calibration. Spectra of different standard stars were
observed for flux calibration.

The reduction of the spectra (bias subtraction, cosmic ray correction,
flat-field correction, 2D-wavelength calibration, night sky subtraction, and
flux calibration) was done in a homogeneous way with IRAF reduction
packages. 
The spectra were not corrected for the variable atmospheric absorption
in the B band.

 Great care was taken to ensure high quality intensity and wavelength
calibrations to keep the intrinsic measurement errors very low.
For a discussion of the intrinsic measurement error, we refer to 
 Gaskell \& Peterson (\cite{gaskell87}). 
 All spectra were calibrated to the same absolute
[\ion{O}{iii}]\,$\lambda$5007 flux of $3.14 \times 10^{-13} \rm erg\,s^{-1}\,cm^{-2}$.
This flux value were
 used in Kollatschny et al. (\cite{kollatschny06}). 
Durret \& Bergeron (\cite{durret88}) and Morris \& Ward (\cite{morris88})
measured similar [\ion{O}{iii}]\,$\lambda$5007 fluxes
in their Mrk\,926 spectra of
$2.05 \times 10^{-13} \rm erg\,s^{-1}\,cm^{-2}$   and 
$3.31 \times 10^{-13} \rm erg\,s^{-1}\,cm^{-2}$, respectively.
%
The accuracy of the [\ion{O}{iii}]\,$\lambda$5007 flux calibration
 was tested for all forbidden emission lines in the spectra.
 We calculated difference spectra for all epochs
 with respect to the mean spectrum of our variability campaign.
Corrections for both small spectral shifts ($<$ 0.5 \AA )
 and small scaling factors were executed
 by minimizing the residuals of the narrow emission lines in the
 difference spectra. 
 All wavelengths were converted to the rest frame of the galaxy (z=0.04701).  
Throughout this paper, we assume that H$_0$~=~70~km s$^{-1}$ Mpc$^{-1}$.
A relative flux  accuracy of better than 1\% was achieved for most of
 our spectra.

\section{Results and discussion}

\subsection{Spectral variations and mean spectra}

\begin{figure*}
 \hbox{
\includegraphics[bb=40 90 380 700,width=9.12cm,angle=270]{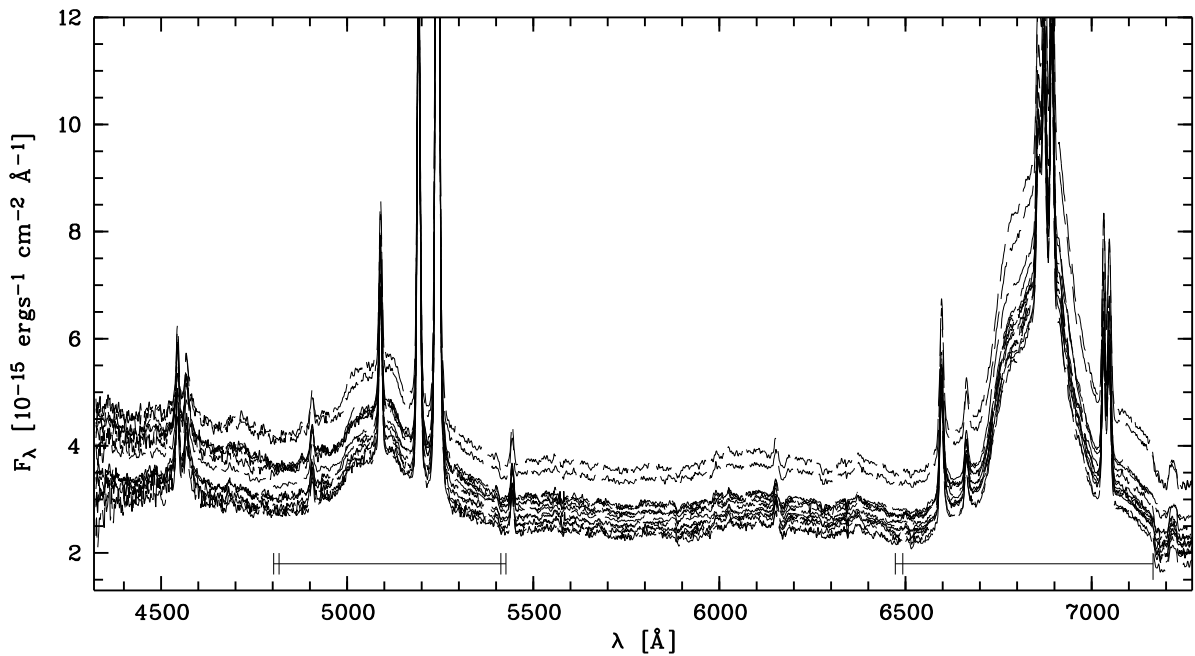}
}
       \vspace*{-2mm} 
  \caption{HET spectra of Mrk\,926 taken between August 3 and
 December 6, 2005. The wavelength ranges for the  H$\beta$ and
 H$\alpha$ lines as well as for the optical continua 
 are indicated at the bottom (see Table ~2).}
   \label{13463fg1.ps}
\end{figure*}

All the optical spectra 
of Mrk\,926 that we obtained during our variability campaign in 2005 are
presented in 
Fig.~\ref{13463fg1.ps}. They show definite variations in the
continuum.
%
%
%
%
   \begin{figure*}
    \includegraphics[bb=40 90 380 700,width=9.12cm,angle=-90]{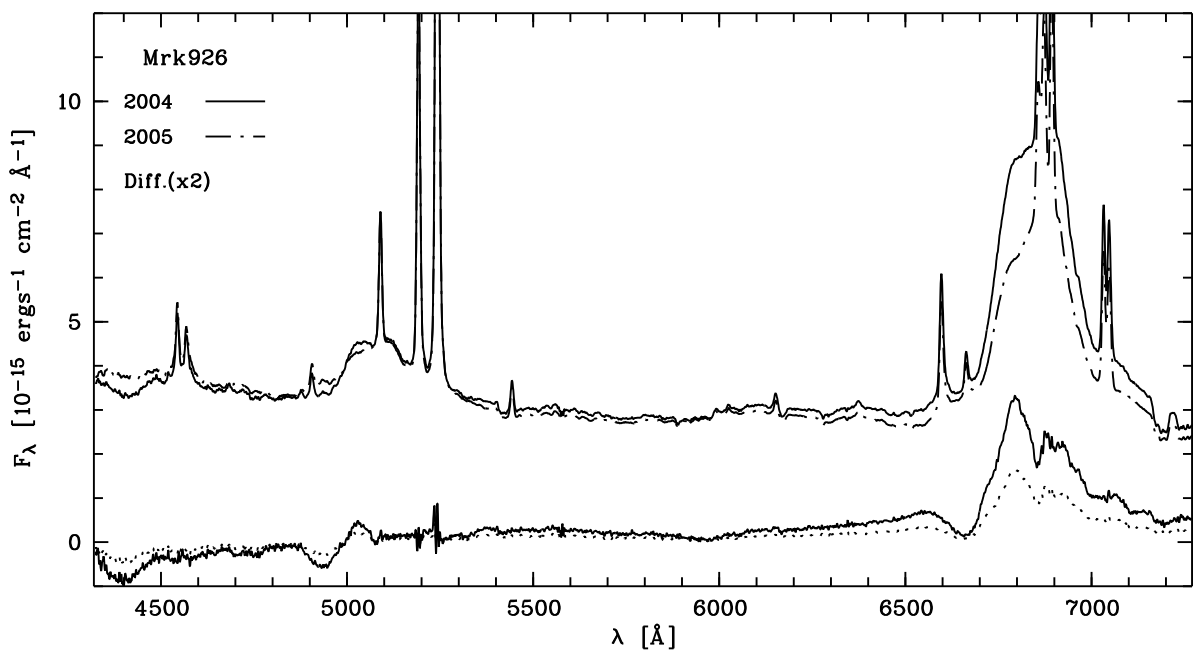}
      \caption{Mean spectra of Mrk\,926
               taken in the years 2004 (solid line) and 2005
               (dash-dotted line).  
              The difference spectrum between the two observing campaigns
              is shown at the bottom (dotted line).  
              It has been scaled by a factor of 2
              (solid line) to enhance the contrast between features.    
              }
         \label{13463fg2.ps}
   \end{figure*}
%
%
   \begin{figure*}
    \includegraphics[bb=40 90 380 700,width=9.12cm,angle=-90]{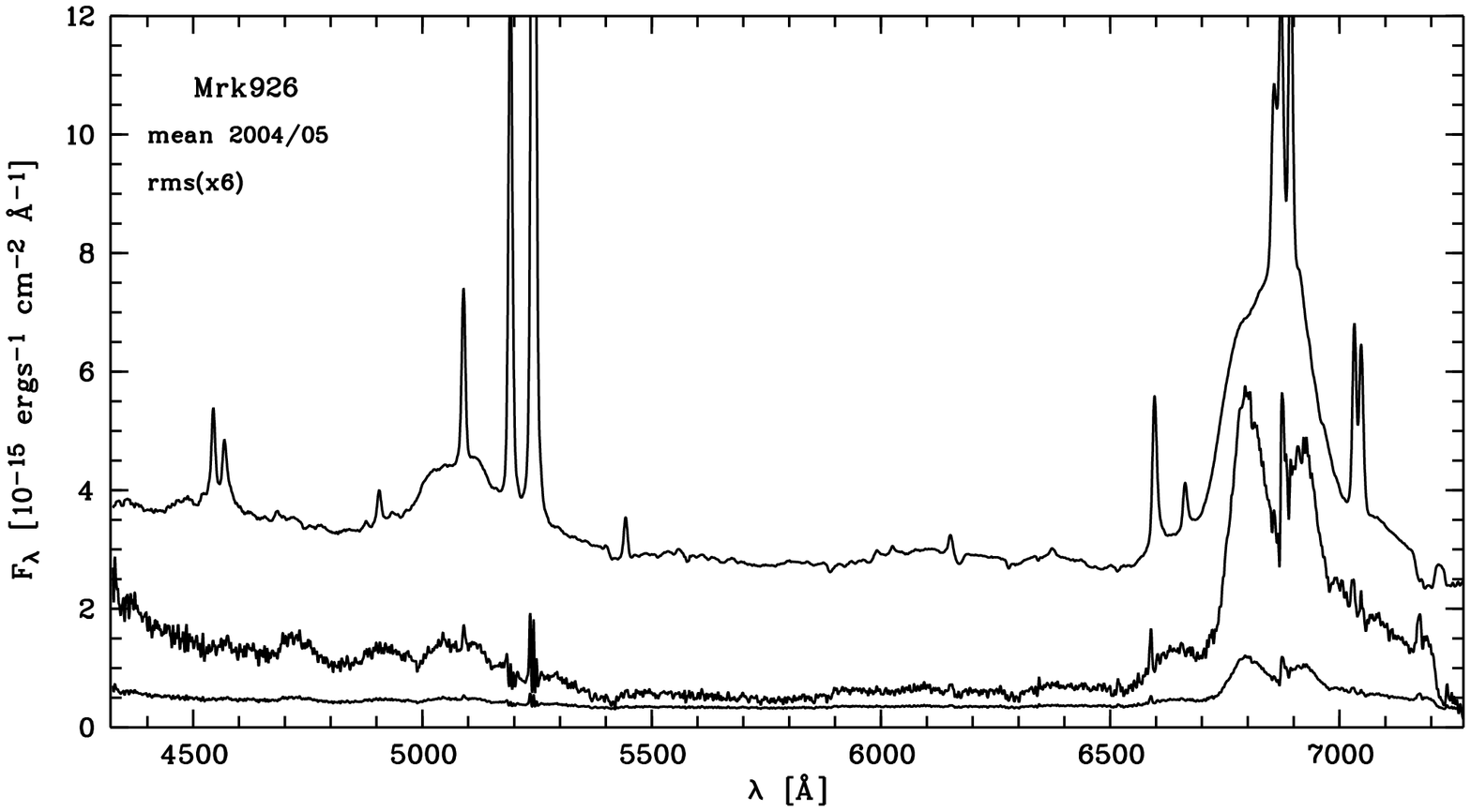}
      \caption{Integrated mean spectrum of Mrk\,926
               for the years 2004 and 2005. The rms spectrum is given at
               the bottom.
               This spectrum has been scaled by a factor
               of 6 (zero level is shifted by -1.5) to enhance
               weaker line structures.
              }
         \label{13463fg3.ps}
   \end{figure*}
%
%
Figure~\ref{13463fg2.ps} shows the mean spectra
of the four epochs taken in the year 2004 (solid line) and of the
 15 epochs taken in 2005
(dash-dotted line).  
The difference spectrum between these two observing years              
is shown at the bottom (dotted line).  
This difference spectrum has in addition been scaled by a factor of 2
(solid line) to show more detailed
small-scale structures in that spectrum.    

Two general results are immediately evident:
the continuum flux was bluer in 2005 than in 2004,
while the mean flux at 6000\,\AA\
was almost equal at both epochs.
The second trend is related to general profile changes in the Balmer lines
H$\gamma$, H$\beta$, and H$\alpha$. The outer blue wings are stronger in
2005 than in 2004 -- seen as absorption troughs in the difference spectra --
while the remaining blue wings as well as 
the red wings show the opposite trend.
This trend is unequally pronounced in the different Balmer lines. 
The H$\alpha$ line shows the strongest variations.
The double peaked H$\alpha$ profile is also clearly distinctive in the
difference spectrum.
The line variations are discussed later in the context of the
line profiles.

Figure~\ref{13463fg3.ps} shows the mean spectrum of Mrk\,926
for both campaigns in 2004 and 2005. The rms spectrum is given at the bottom
with two different vertical scalings.
It has in addition been scaled by a factor of 6 
(zero level is shifted by -1.5) to enhance weaker line structures.
The rms spectra show the variable parts of the line profiles.
Besides an inner double-peaked profile - as seen in the difference H$\alpha$
spectrum - two separate outer blue and red components are visible
in the H$\alpha$ and H$\beta$ rms spectra.
These spectra are also discussed later in the context of the
line profiles.

We inspected the spectra for continuum regions that are
free of both strong emission and absorption lines.
Because of the extreme line widths of the broad emission line profiles
in  Mrk\,926, it is 
difficult to confine these regions. We finally 
identified four continuum regions close to 4600, 5180,
5500, and 6200\AA{} (see Table~2),
\setcounter{table}{1}
\begin{table}
\centering
\tabcolsep+4.5mm
\caption{Boundaries of mean continuum values and line integration limits}
\begin{tabular}{lccccccc}
\hline 
\noalign{\smallskip}
Cont./Line & Wavelength range & Pseudo-continuum \\
\noalign{\smallskip}
(1) & (2) & (3) \\
\noalign{\smallskip}
\hline 
\noalign{\smallskip}
Cont.~4600         & 4587\,\AA\ -- 4601\,\AA \\
Cont.~5180         & 5170\,\AA\ -- 5183\,\AA \\
Cont.~5500         & 5472\,\AA\ -- 5513\,\AA \\
Cont.~6200         & 6182\,\AA\ -- 6201\,\AA \\
H$\beta$           & 4601\,\AA\ -- 5170\,\AA & 4600\,\AA\ -- 5180\,\AA \\
H$\alpha$          & 6201\,\AA\ -- 6843\,\AA & 6200\,\AA\ \\
\noalign{\smallskip}
\hline \\
\end{tabular}
\end{table}
of widths between 15 and 40\,\AA{}.  

\subsection{Continuum and emission-line light curves}

It is important to identify wavelength ranges
in AGN spectra that are free of both emission and absorption lines 
to derive their continuum light curves.
The continuum region at 5100 \AA\ has widely been used
in studies of many AGNs, such as 
the prototype Seyfert galaxy NGC\,5548 (Peterson et al. \cite{peterson92}).
The spectral range between 5130 and 5140 \AA\ was found
to be a more appropriate continuum region  
 in the narrow-line Seyfert galaxy Mrk\,110
(Kollatschny et al. \cite{kollatschny03}).
However, because of the extreme H$\beta$ line width (see Fig.~2) of Mrk\,926
the most appropriate continuum section 
 is the wavelength interval 5170 -- 5183\,\AA{}. 
Three additional optical continuum ranges were selected close to
4600\AA{}, 5500\AA{}, and 6200\AA{} (see Table~2).

We integrated the
broad emission-line intensities of H$\beta$ and H$\alpha$,
between the boundaries listed in Table~2.
Figure~1 shows the selected
wavelength ranges for the  H$\beta$ and
H$\alpha$ lines as well as for their related optical continuum ranges.

We first
subtracted a linear pseudo-continuum defined by the boundaries
given in Table~2 . 
For the H$\alpha$ line, 
we found only one continuum data point only 
on the blue side of this line.
For our intensity measurements, we assumed that the continuum
is flat below H$\alpha$. 
Constant narrow line components were
subtracted first before we measured the broad line intensities.
The results of the continuum and line intensity measurements are given in 
Table\,3. 
\begin{table}
\tabcolsep+0.5mm
\caption{Continuum and integrated line fluxes}
\begin{tabular}{ccccc}
\noalign{\smallskip}
\hline 
\noalign{\smallskip}
Julian Date & 4600\,\AA & 5180\,\AA  & H$\beta$ & H$\alpha$ \\ 
2\,400\,000+\\
(1) & (2) & (3) & (4) & (5) \\ 
\noalign{\smallskip}
\hline
\noalign{\smallskip}
53224.362 & 3.294 $\pm\ $ 0.066 & 2.942 $\pm\ $ 0.059 & 373.4  $\pm\ $ 7.5 & 1453. $\pm\ $ 31. \\
53238.332 & 3.246 $\pm\ $ 0.065 & 2.899 $\pm\ $ 0.058 & 391.3  $\pm\ $ 7.8 & 1449. $\pm\ $ 29. \\
53258.283 & 3.426 $\pm\ $ 0.068 & 3.178 $\pm\ $ 0.064 & 401.6  $\pm\ $ 8.0 & 1464. $\pm\ $ 29. \\
53287.194 & 2.976 $\pm\ $ 0.060 & 2.689 $\pm\ $ 0.054 & 392.8  $\pm\ $ 7.8 & 1460. $\pm\ $ 29. \\
53585.380 & 3.029 $\pm\ $ 0.061 & 2.801 $\pm\ $ 0.056 & 387.6  $\pm\ $ 7.7 & 1297. $\pm\ $ 26. \\
53590.362 & 2.710 $\pm\ $ 0.054 & 2.478 $\pm\ $ 0.050 & 333.6  $\pm\ $ 6.7 & 1180. $\pm\ $ 24. \\
53600.329 & 2.766 $\pm\ $ 0.055 & 2.407 $\pm\ $ 0.048 & 341.0  $\pm\ $ 6.8 & 1185. $\pm\ $ 24. \\
53609.318 & 3.002 $\pm\ $ 0.060 & 2.653 $\pm\ $ 0.053 & 369.9  $\pm\ $ 7.4 & 1245. $\pm\ $ 25. \\
53628.262 & 2.840 $\pm\ $ 0.057 & 2.494 $\pm\ $ 0.050 & 372.2  $\pm\ $ 7.4 & 1231. $\pm\ $ 25. \\
53629.255 & 2.767 $\pm\ $ 0.083 & 2.392 $\pm\ $ 0.096 & 354.2  $\pm\ $14.2 & 1102. $\pm\ $ 44. \\
53644.212 & 3.477 $\pm\ $ 0.069 & 2.964 $\pm\ $ 0.059 & 422.1  $\pm\ $ 8.4 & 1206. $\pm\ $ 24. \\
53653.194 & 2.975 $\pm\ $ 0.060 & 2.606 $\pm\ $ 0.052 & 363.5  $\pm\ $ 7.3 & 1160. $\pm\ $ 23. \\
53670.147 & 3.232 $\pm\ $ 0.065 & 2.780 $\pm\ $ 0.056 & 387.1  $\pm\ $ 7.7 & 1111. $\pm\ $ 22. \\
53674.144 & 3.565 $\pm\ $ 0.071 & 3.030 $\pm\ $ 0.061 & 405.4  $\pm\ $ 8.1 & 1235. $\pm\ $ 25. \\
53680.109 & 3.602 $\pm\ $ 0.072 & 2.894 $\pm\ $ 0.058 & 374.6  $\pm\ $ 7.5 & 1167. $\pm\ $ 23. \\
53686.096 & 3.572 $\pm\ $ 0.071 & 2.976 $\pm\ $ 0.059 & 381.1  $\pm\ $ 7.6 & 1239. $\pm\ $ 25. \\
53693.090 & 4.060 $\pm\ $ 0.081 & 3.400 $\pm\ $ 0.068 & 456.0  $\pm\ $ 9.1 & 1333. $\pm\ $ 27. \\
53700.070 & 4.140 $\pm\ $ 0.082 & 3.606 $\pm\ $ 0.072 & 474.1  $\pm\ $ 9.5 & 1413. $\pm\ $ 28. \\
53710.040 & 3.646 $\pm\ $ 0.073 & 2.868 $\pm\ $ 0.057 & 383.2  $\pm\ $ 7.7 & 1199. $\pm\ $ 24. \\
\noalign{\smallskip}
\hline 
\noalign{\smallskip}
\end{tabular}

Continuum fluxes (2) - (5) in
10$^{-15}$\,erg\,s$^{-1}$\,cm$^{-2}$\,\AA$^{-1}$.\\
Line fluxes (6) - (9) in 10$^{-15}$\,erg\,s$^{-1}$\,cm$^{-2}$.
\end{table}

We derived
 a mean continuum flux of
$F_{\lambda}(5180$\,\AA$)$ = $2.82\pm0.23 \times 
10^{-15}$\,erg\,s$^{-1}$\,cm$^{-2}$\,\AA$^{-1}$
(from Table~3) at 5180$\,\AA$  for 2005.
 This corresponds to a mean luminosity of
$L(5180$\,\AA$) = 1.159\pm0.094 \times 10^{40}$\,erg\,s$^{-1}$\,\AA$^{-1}$
or
$\lambda$\,$L_{\lambda}(5180$\,\AA$) = 5.91\pm0.56 \times
10^{43}$\,erg\,s$^{-1}$. 
The mean H$\beta$ flux amounts to 
F(H$\beta)$ = 3.87$\pm0.39 \times 10^{-13}$\,erg\,s$^{-1}$\,cm$^{-2}$,
and the mean H$\beta$ luminosity to
L(H$\beta$) = 1.59$\pm0.16 \times 10^{42}$\,erg\,s$^{-1}$\,\AA$^{-1}$.

In Figure~4  we present the light curves of the continuum fluxes at 4600 and
 5180\,\AA\
 (in units of 10$^{-15}$ erg cm$^{-2}$ s$^{-1}$\,\AA$^{-1}$) and
the light curves of the integrated
  emission-line fluxes of H$\beta$ and H$\alpha$
 (in units of  10$^{-15}$ erg cm$^{-2}$ s$^{-1}$)
for both campaigns performed in 2004 and 2005 
and in greater detail for 2005 (right column).
Conspicuous variability amplitudes in the
continuum as well as
in the emission line fluxes
are to be seen on timescales of a few days only.

%
\begin{figure*}
\hbox{\includegraphics[bb=40 90 380 700,width=55mm,height=85mm,angle=270]{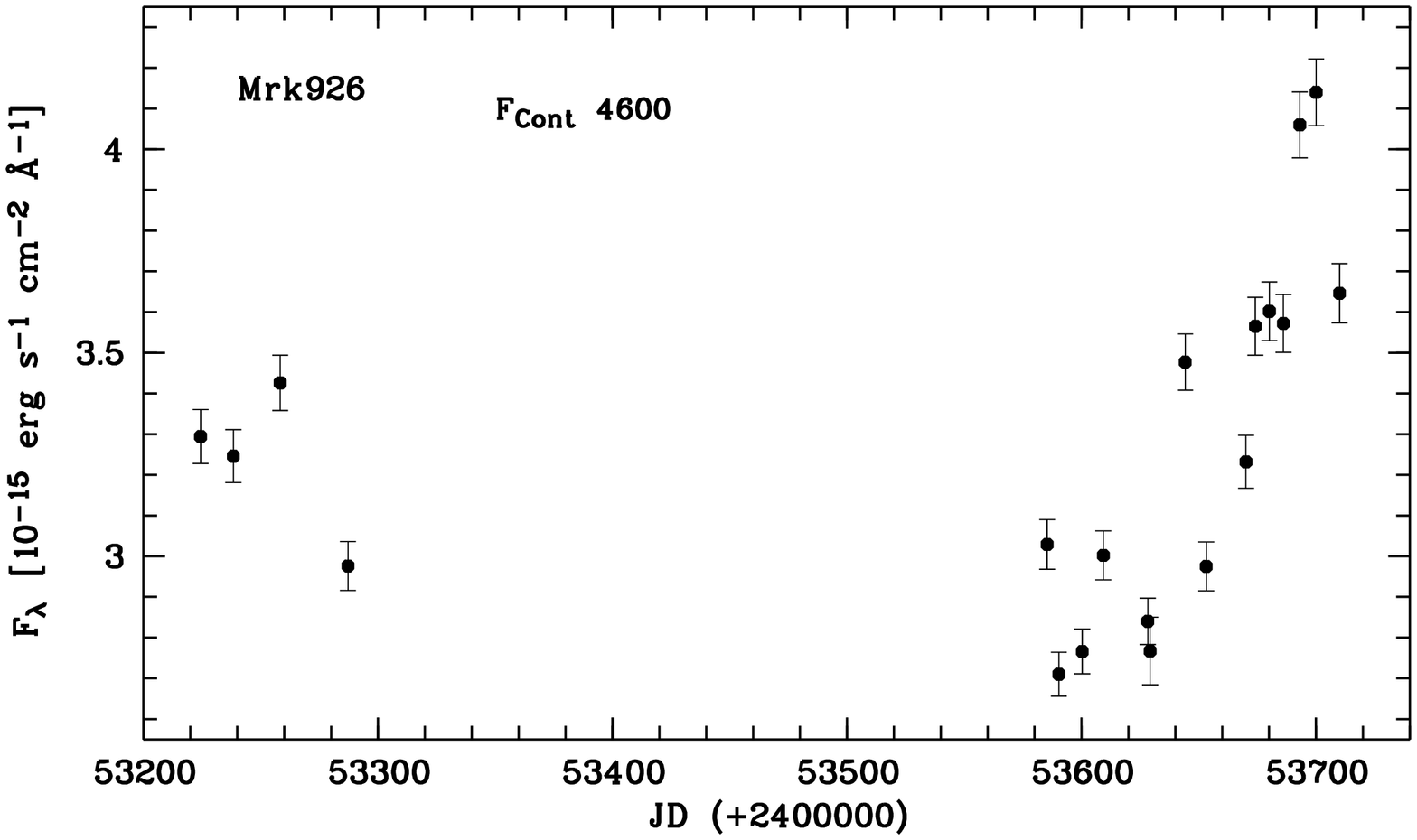}\hspace*{7mm}      
\includegraphics[bb=40 90 380 700,width=55mm,height=85mm,angle=270]{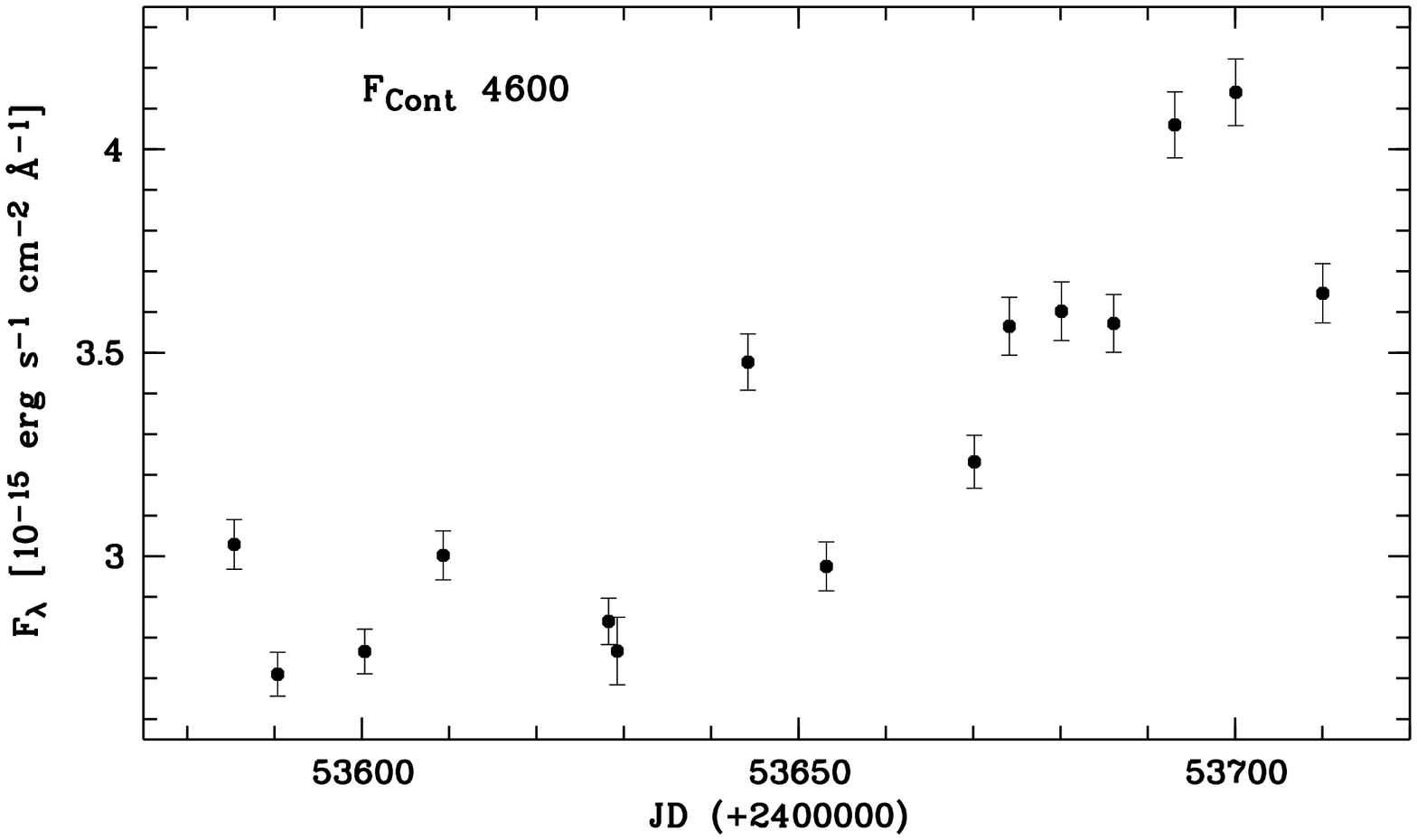}}
\hbox{\includegraphics[bb=40 90 380 700,width=55mm,height=85mm,angle=270]{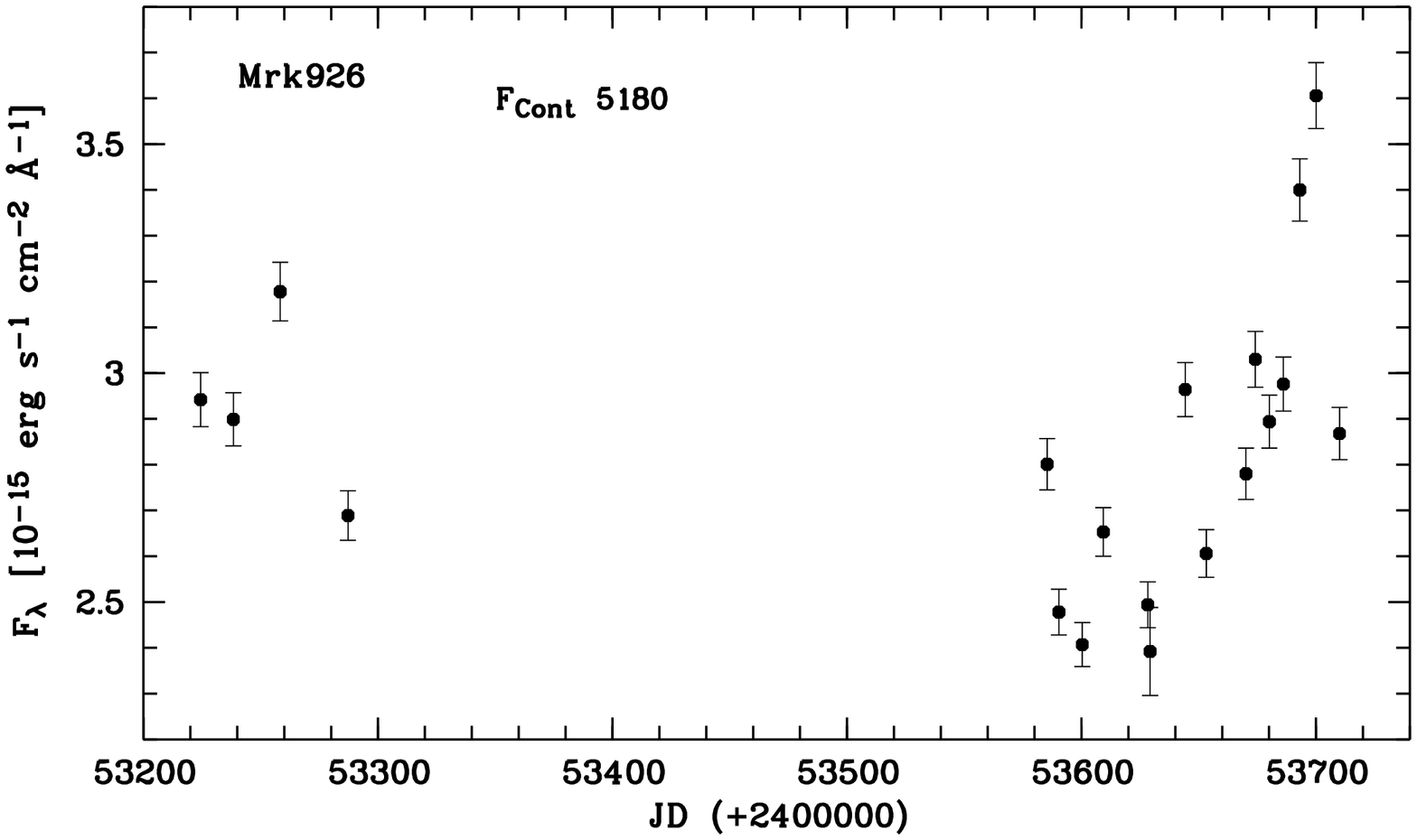}\hspace*{7mm}      
\includegraphics[bb=40 90 380 700,width=55mm,height=85mm,angle=270]{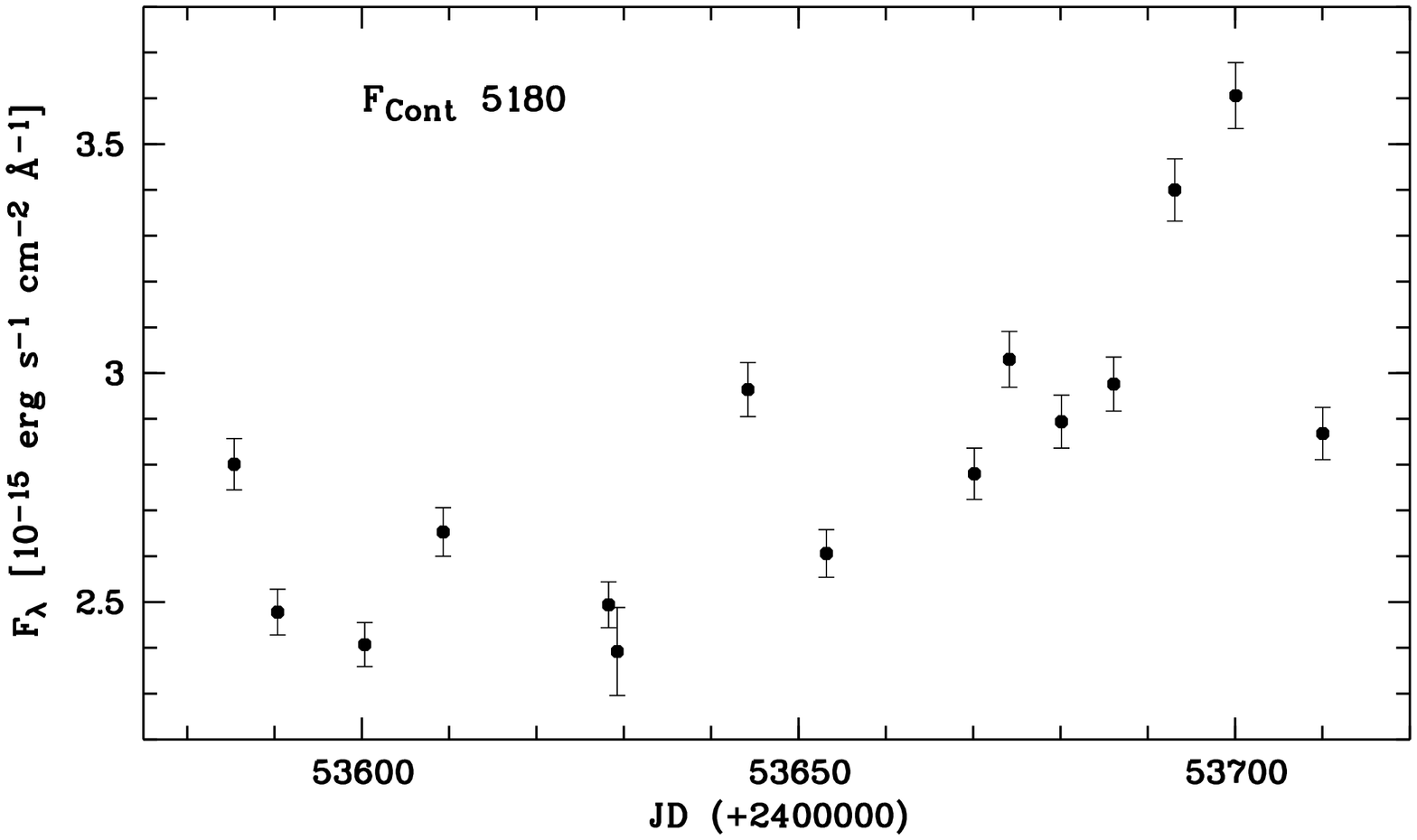}}
\hbox{\includegraphics[bb=40 90 380 700,width=55mm,height=85mm,angle=270]{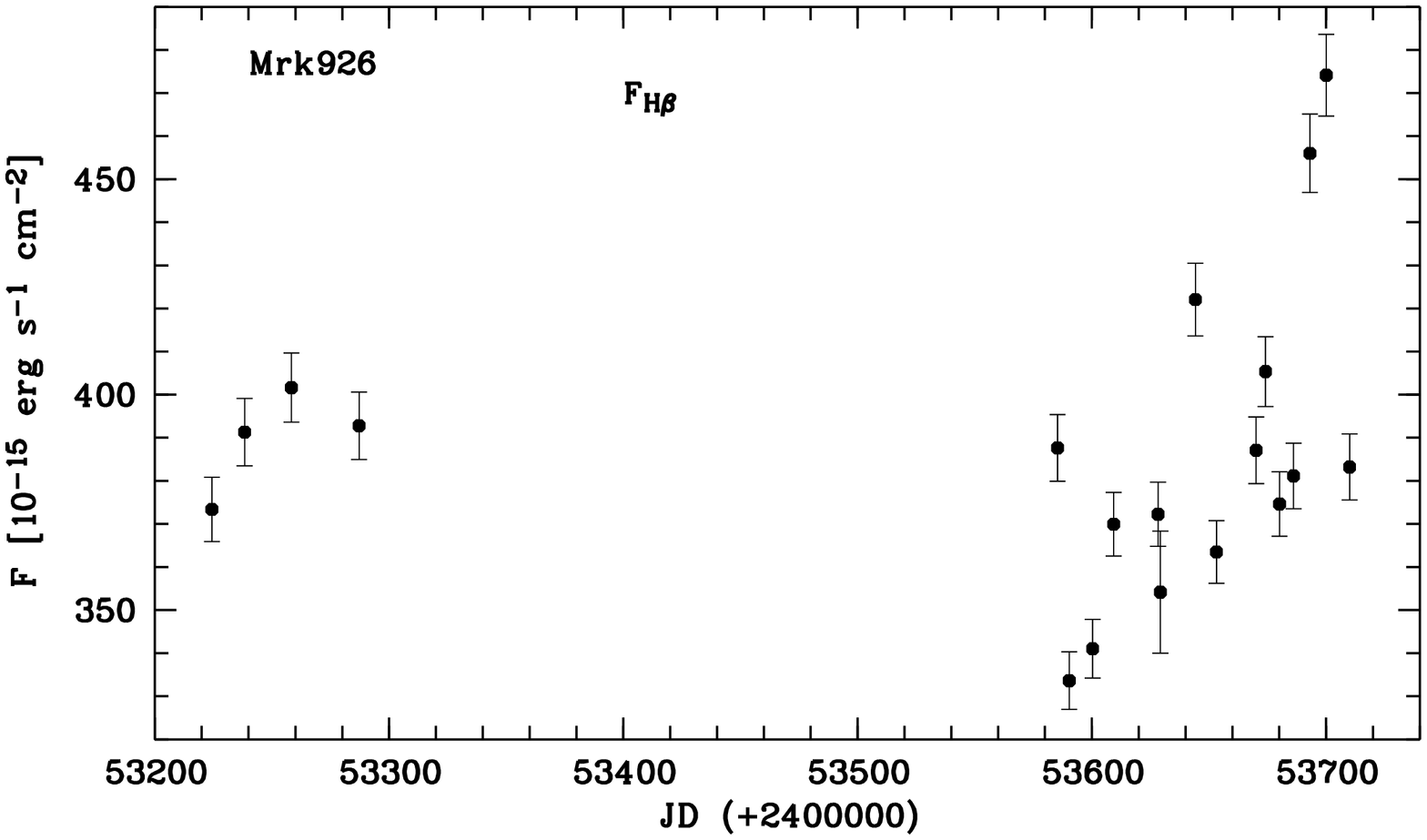}\hspace*{7mm}     
\includegraphics[bb=40 90 380 700,width=55mm,height=85mm,angle=270]{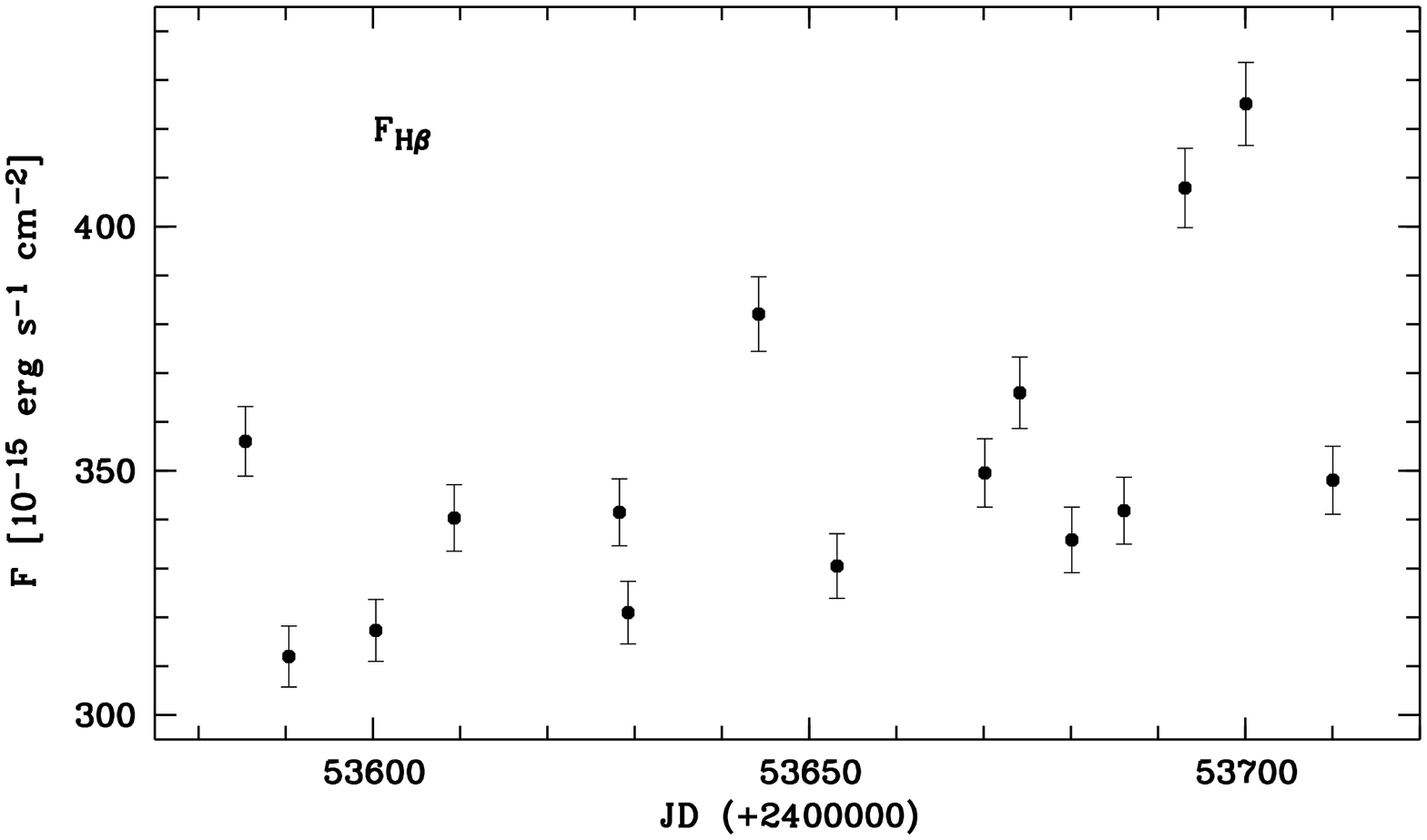}}
\hbox{\includegraphics[bb=40 90 380 700,width=55mm,height=85mm,angle=270]{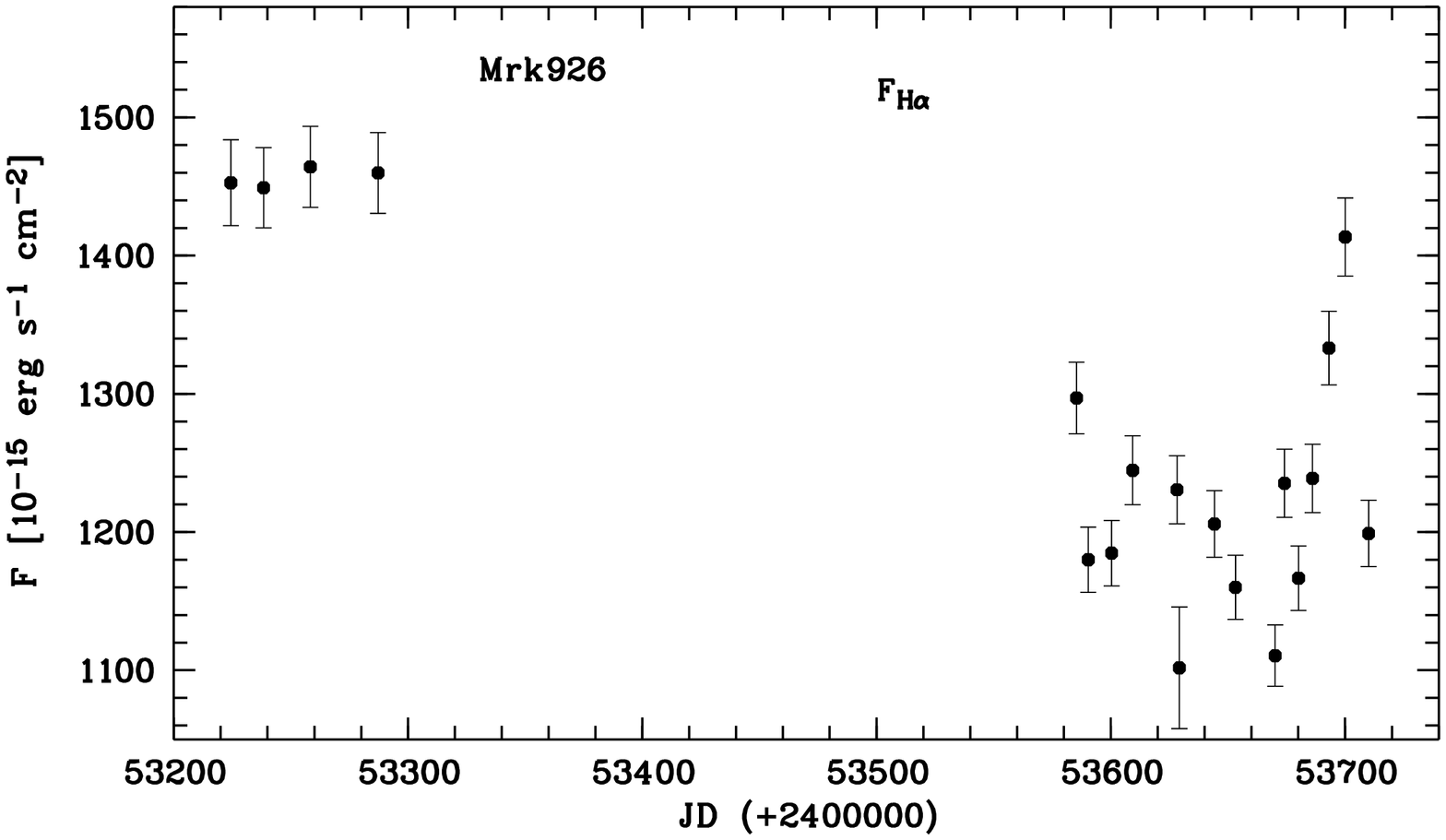}\hspace*{7mm}
\includegraphics[bb=40 90 380 700,width=55mm,height=85mm,angle=270]{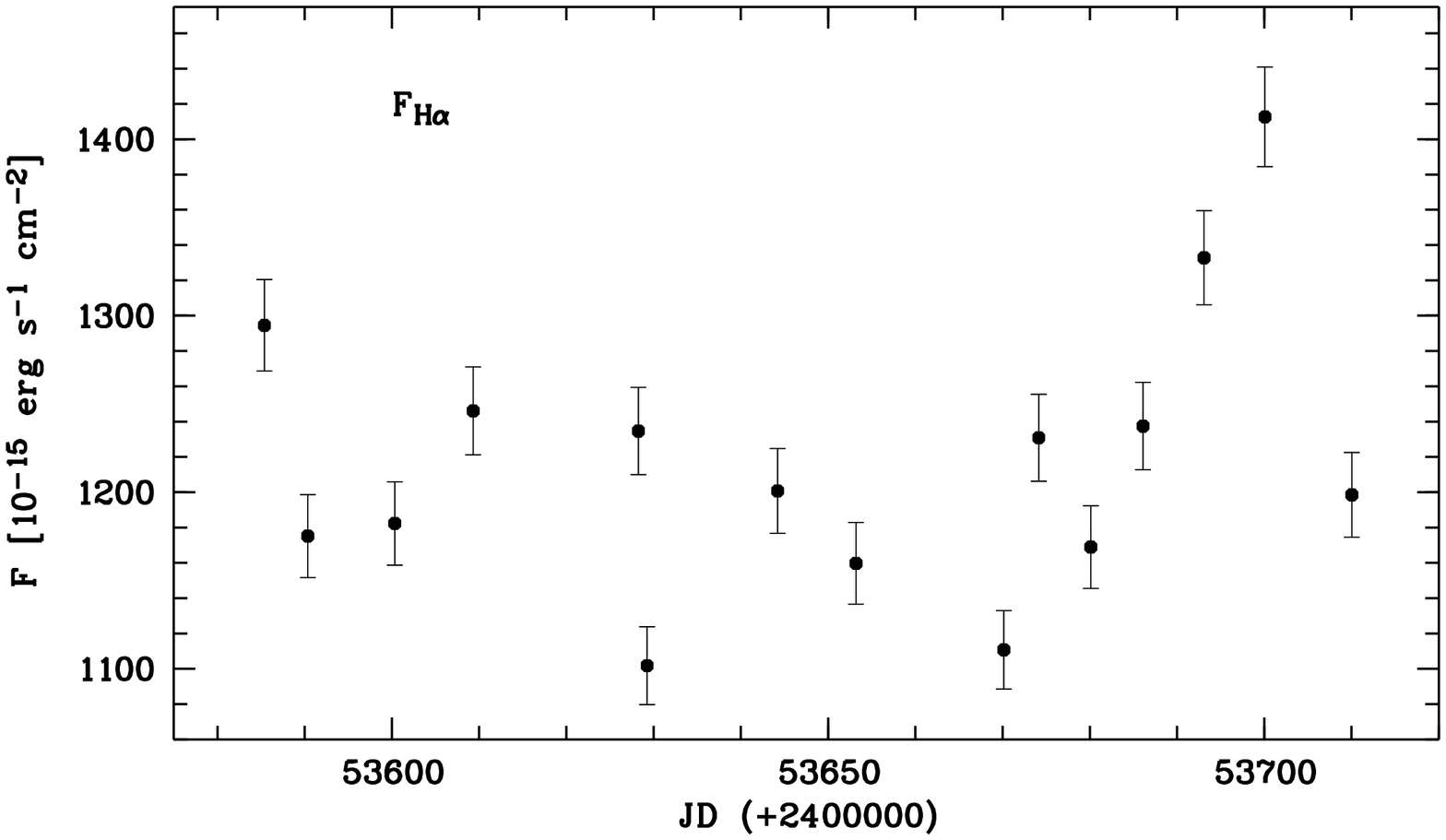}}
       \vspace*{5mm} 
  \caption {Light curves of the continuum fluxes at 4600 and 5180\,\AA\
 (in units of 10$^{-15}$ erg cm$^{-2}$ s$^{-1}$\,\AA$^{-1}$) and of
the integrated emission line fluxes of H$\beta$ and H$\alpha$
   (in units of  10$^{-15}$ erg cm$^{-2}$ s$^{-1}$) for the years 2004 and 2005 and in greater detail for the
 year 2005 (right column).
}
\end{figure*}
%
%
%
We now provide some
statistics of the continuum and emission line intensity variations
in Table\,4.
We indicate the minimum and maximum fluxes F$_{min}$ and F$_{max}$,
peak-to-peak amplitudes R$_{max}$ = F$_{max}$/F$_{min}$, the mean flux
over the period of observations $<$F$>$, the standard deviation $\sigma_F$,
 and the fractional variation 
\[ F_{var} = \frac{\sqrt{{\sigma_F}^2 - \Delta^2}}{<F>} \] 
as defined by Rodr\'\i{}guez-Pascual et al.\ (\cite{rodriguez97}).
\begin{table}
\centering
\tabcolsep+2.5mm
\caption{Variability statistics for Mrk\,926 in the year 2005 (upper half)
 as well as for both years 2004/2005 (lower half)}
\begin{tabular}{lccccccc}
\hline 
\noalign{\smallskip}
Cont./Line & F$_{min}$ & F$_{max}$ & R$_{max}$ & $<$F$>$ & $\sigma_F$ & F$_{var
}$ \\
\noalign{\smallskip}
(1) & (2) & (3) & (4) & (5) & (6) & (7) \\ 
\noalign{\smallskip}
\hline 
\noalign{\smallskip}
Cont.~4600         & 2.71 &  4.14  & 1.53 & 3.29  & 0.468 & 0.141 \\
Cont.~5180         & 2.39 &  3.61  & 1.51 & 2.82  & 0.349 & 0.122 \\
H$\beta$           & 333.6 & 474.1 & 1.42 & 387.1 & 38.93 & 0.098 \\
H$\alpha$          & 1102. & 1413. & 1.28 & 1220. & 81.66 & 0.063 \\
\noalign{\smallskip}
\hline 
\noalign{\smallskip}
Cont.~4600         & 2.71 &  4.14  & 1.53 & 3.28  & 0.420 & 0.117 \\
Cont.~5180         & 2.39 &  3.61  & 1.51 & 2.85  & 0.320 & 0.111 \\
H$\beta$           & 333.6 & 474.1 & 1.42 & 387.6 & 34.69 & 0.087 \\
H$\alpha$          & 1102. & 1534. & 1.39 & 1273. & 128.9 & 0.099 \\
\noalign{\smallskip}
\hline 
\end{tabular}
\end{table}
The intrinsic variations are of the order of 7 to 15\%.

\subsection{Difference line profiles}

\begin{figure}
 \includegraphics[bb=40 60 570 780,width=63mm,angle=270]{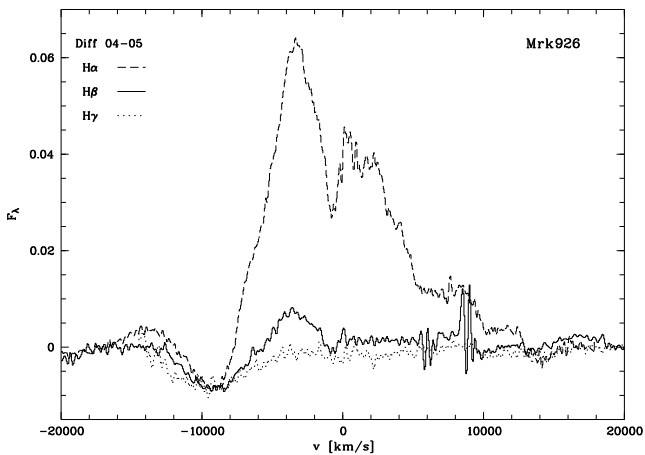}
  \caption{Difference line profiles of H$\alpha$ (dashed line), H$\beta$
 (solid line), and  H$\gamma$ (dotted line) in velocity space for
 the two campaigns in 2004 and 2005.}
  \label{13463fg5.ps}
\end{figure}

Figure~\ref{13463fg5.ps} shows the difference line profiles of
H$\alpha$ (dashed line), H$\beta$ (solid line), and  H$\gamma$ (dotted line)
in velocity space
for the two observing campaigns in the years 2004 and 2005.
%
%

%

The difference spectra of H$\alpha$, H$\beta$, and H$\gamma$ have common 
characteristics. An additional intensity trend is
 superimposed decreasing from
H$\alpha$ to H$\gamma$ : both H$\alpha$ and
 H$\beta$ show double-peaked profiles
with two separate broad components centered at -3400$\pm$ 200 and
-1500$\pm$ 200 km s$^{-1}$. These two components are not located symmetrically 
with respect to v=0 km s$^{-1}$.
The blue line component is stronger than the red one indicating
that there are stronger
variations in the blue rather than the red wing.
Furthermore,
all three difference spectra show a broad minimum of identical intensity
at -9000$\pm$ 500 km s$^{-1}$. This is the only
section in all Balmer line profiles where the 2004 spectra were more intense
than the 2005 spectra.

The relative strengths of the difference line
intensities with respect to their corresponding mean line intensities
are different for the three Balmer lines (see Fig.~\ref{13463fg2.ps}).
The intensity of the H$\alpha$ difference spectrum
(between v=-17~000,+19~000 km s$^{-1}$)
corresponds to 20\% of the H$\alpha$ mean spectrum. 
However, the intensity of the H$\beta$ difference spectrum
is small compared to the mean
H$\beta$ spectrum, i.e., 0.5\% or rather 5\% when considering the positive
fraction of the difference spectrum only.
\subsection{Mean line profiles}

\begin{figure}
 \includegraphics[bb=40 60 570 780,width=63mm,angle=270]{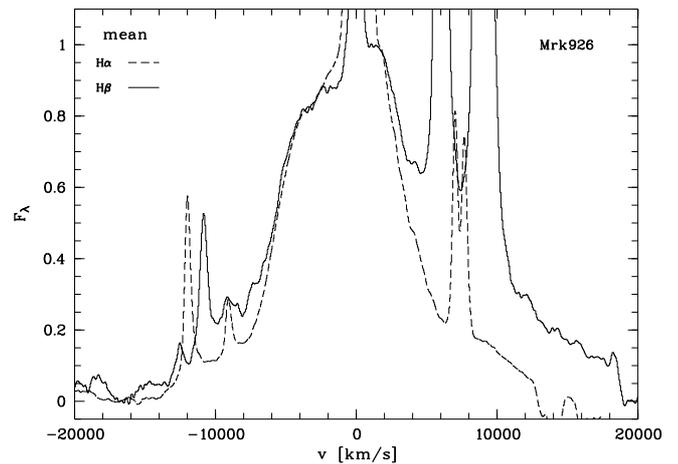}
  \caption{Normalized mean line profiles of
    H$\alpha$ (dashed line) and  H$\beta$ (solid line).}
  \label{13463fg6.ps}
\end{figure}

Normalized mean and rms
line profiles of H$\alpha$ and
 H$\beta$ are presented in Figs.\,6 to 10
in velocity space.
The mean H$\alpha$ and H$\beta$ profiles are free of any
detailed structures (Fig.~\ref{13463fg6.ps}).
%
%
\begin{figure}
 \includegraphics[bb=40 60 570 780,width=63mm,angle=270]{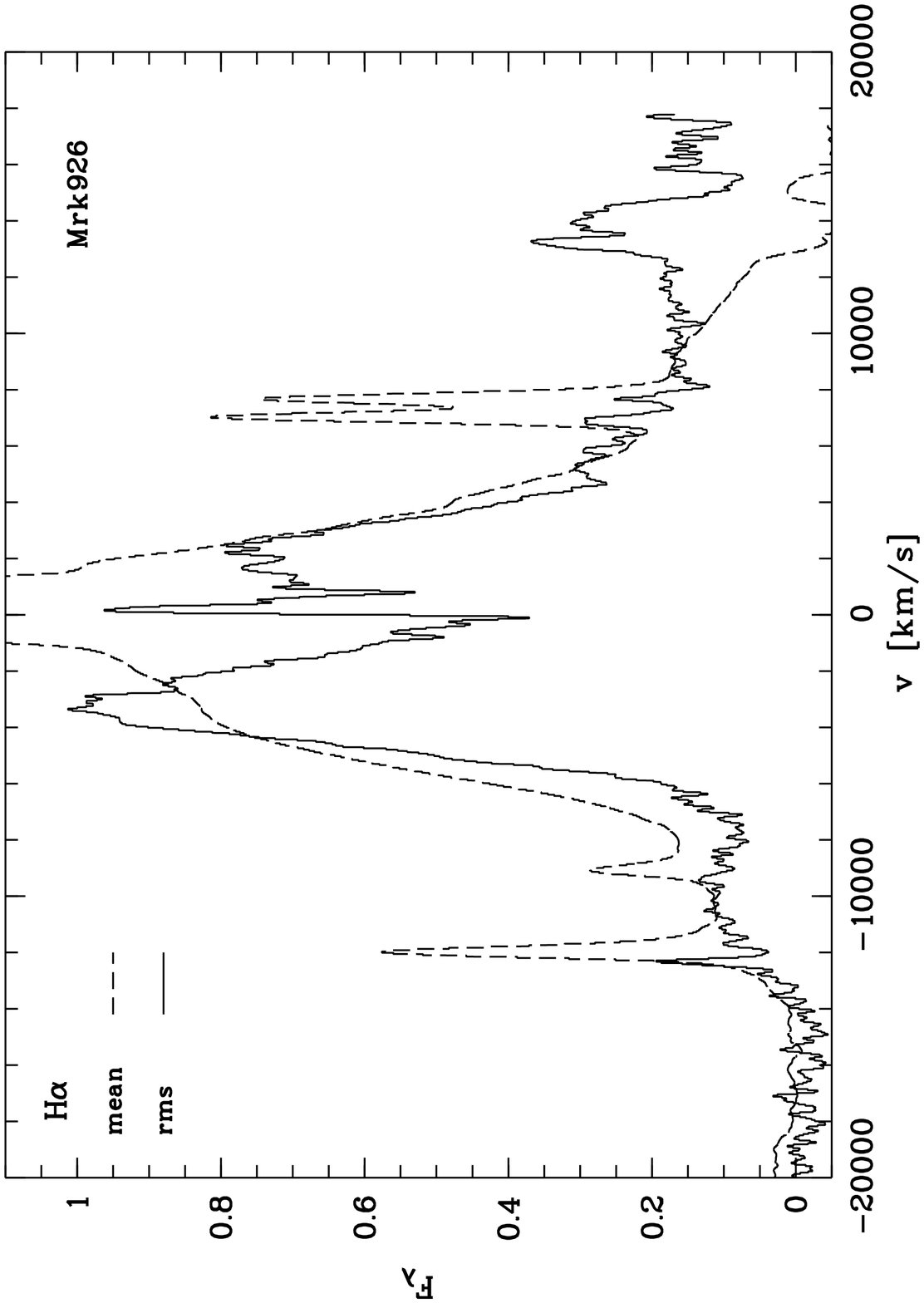}
  \caption{Normalized mean (dashed line) and rms (solid line) line
 profiles of H$\alpha$
   in velocity space.}
  \label{13463fg7.ps}
\end{figure}
\begin{figure}
  \includegraphics[bb=40 60 570 780,width=63mm,angle=270]{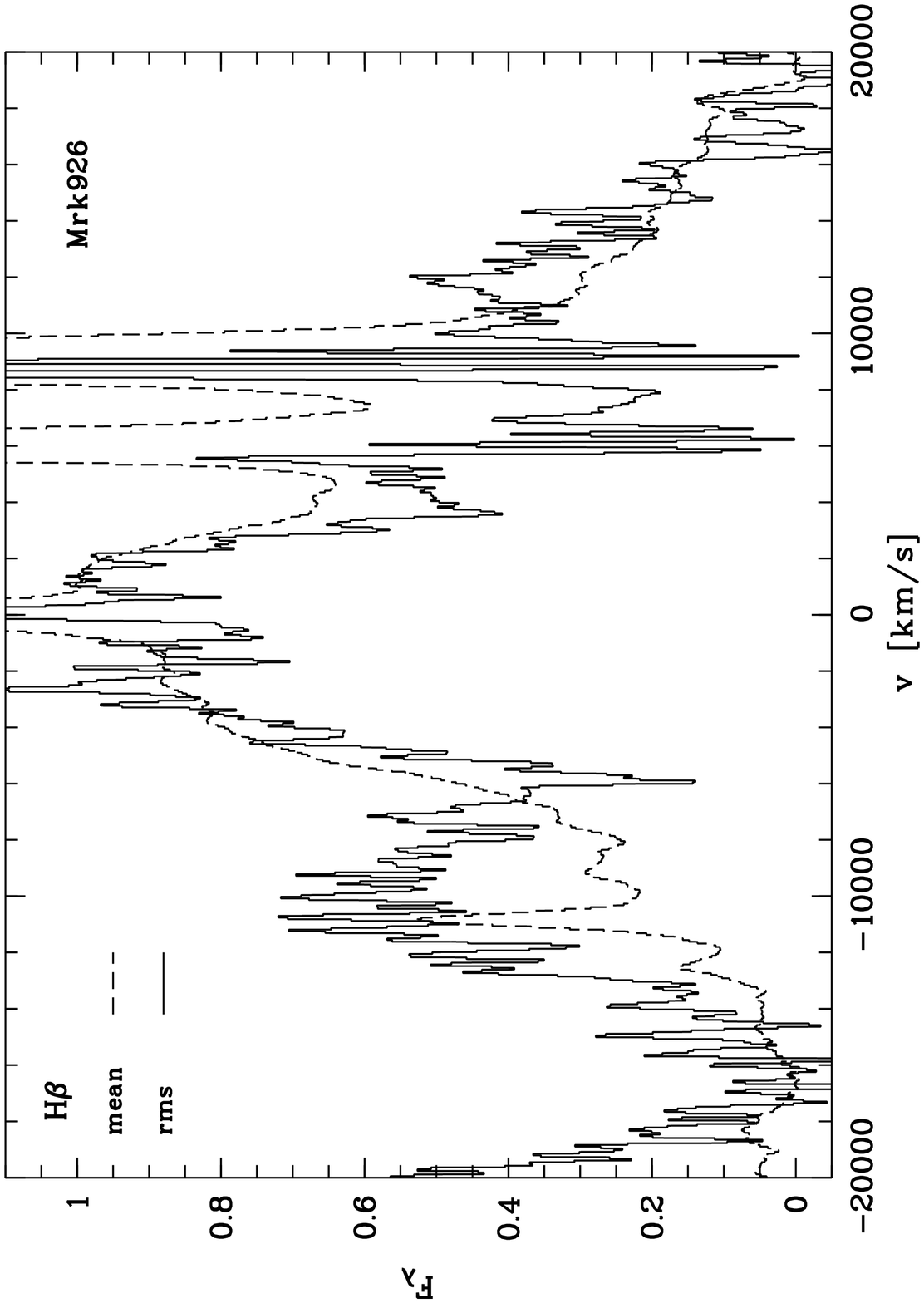}
  \caption{Normalized mean (dashed line) and rms (solid line)
 line profiles of H$\beta$ in velocity space.}
  \label{13463fg8.ps}
\end{figure}
\begin{figure}
 \includegraphics[bb=40 60 570 780,width=63mm,angle=270]{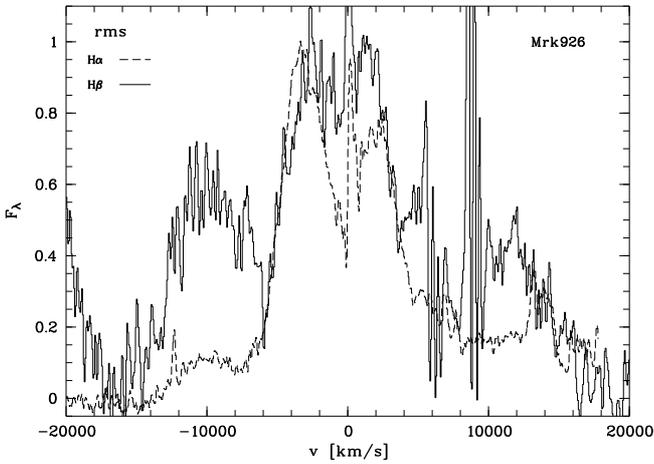}
  \caption{Normalized rms line profiles of
    H$\alpha$ (dashed line) and  H$\beta$ (solid line)
   in velocity space.}
  \label{13463fg9.ps}
\end{figure}
\begin{figure}
 \includegraphics[bb=40 60 570 780,width=63mm,angle=270]{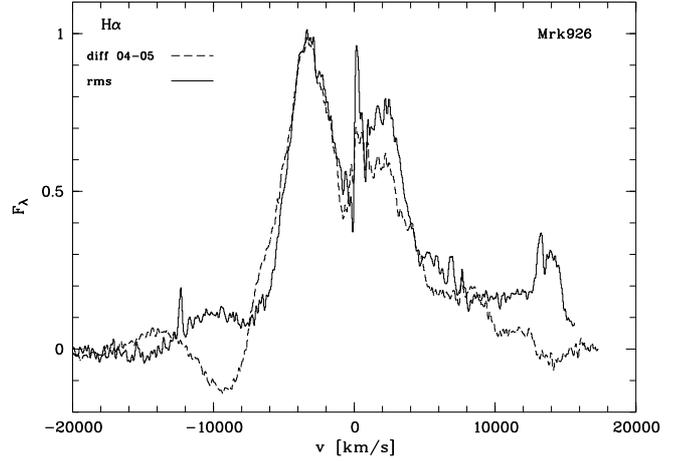}
  \caption{Normalized rms (solid line) and difference (dashed line)
  line profiles of H$\alpha$ in velocity space.}
  \label{13463fg10.ps}
\end{figure}
\subsection{Rms line profiles}
The rms spectra illustrate the variations in the line profile segments
during our variability campaign in the
years 2004 and 2005. 
Our rms profiles
(Figs.~\ref{13463fg7.ps}, \ref{13463fg8.ps},
\ref{13463fg9.ps}) consist of five distinguishable components:  
a narrow central component, two broad inner blue and red wing components,
and two even broader outer wing components.
The intervals of the individual rms line segments are given in Table~5.
\begin{table}
\centering
\tabcolsep+2.5mm
\caption{Intervals of the H$\beta$ and H$\alpha$ line-profile segments.}
\begin{tabular}{l D{t}{\,\textrm{to}\,}{7,7}D{t}{\,\textrm{to}\,}{7,7}}
\hline 
\noalign{\smallskip}
Segment & \multicolumn{1}{c}{H$\beta$} & \multicolumn{1}{c}{H$\alpha$}\\
     & \multicolumn{1}{c}{[km s$^{-1}$]} & \multicolumn{1}{c}{[km s$^{-1}$]} \\
 \multicolumn{1}{c}{(1)}  &  \multicolumn{1}{c}{(2)} &  \multicolumn{1}{c}{(3)} \\ 
\noalign{\smallskip}
\hline
\noalign{\smallskip}
outer blue & -16~000 t - 6,000 & -14~000 t - 7~800\\
inner blue &  -6~000 t - 220   &  -7~800 t - 100\\
central    &   -200 t  ~~600     &    -100 t ~~800\\ 
inner red  &   600 t ~~7~500    &   800 t ~~7~500\\ 
outer red  &   ~7~500 t ~~19~000 &  ~7,500 t ~>16~000\\
\noalign{\smallskip}
\hline 
\end{tabular}
\end{table}
%
%
The uncertainty in the dividing line of the intervals
amounts to 200 km s$^{-1}$ for the central component and
300 km s$^{-1}$ for the inner components.
However, the separation between the inner and outer red component is
quite  uncertain
because of the H$\alpha$ line blending with the [SII]-lines
and the H$\beta$ line blending with
the [OIII]-lines, the uncertainty being as large as ~3~000  
km s$^{-1}$. The red end of the H$\alpha$ line profile is beyond our observed
spectral range.
The central component and the outer rms components are not arranged
strictly symmetrical 
with respect to v = 0~km s$^{-1}$.

The outer wing components in H$\alpha$ and H$\beta$ 
are broader than the inner wing components.
The relative intensities of the outer components with respect
to the inner
ones are different for the H$\alpha$ and H$\beta$ lines, i.e.,
the outer blue component
is far stronger in H$\beta$ than in  H$\alpha$
(Fig.~\ref{13463fg9.ps}). This causes a shift in the dividing line 
between outer and inner blue component in H$\beta$.
Therefore, at least some of the different line segment
 limits in  H$\alpha$ and
H$\beta$ might be caused by the relative intensities
of the line segments.
The maxima of the
outer blue
components of H$\alpha$ and H$\beta$ are at
-9~500 km s$^{-1}$  in both profiles. 
One may suspect that the strong outer blue component in the
H$\beta$ rms profile is caused
by a blend with a variable underlying \ion{He}{ii}\,$\lambda 4686$ line
(see Fig.~\ref{13463fg10.ps}),
as seen in Mrk\,110 (Kollatschny et al.
\cite{kollatschny01}).
However, this is less likely to be true for Mrk\,926:
the \ion{He}{ii}\,$\lambda 4686$ line is centered on -10~800 km s$^{-1}$
but the observed outer blue wing is centered  on -9~500 km s$^{-1}$
(Fig.~\ref{13463fg8.ps}).
 
The blue-to-red intensity ratios of the inner line components
in the rms profiles
are different for
 H$\alpha$ and  H$\beta$ (Fig.~\ref{13463fg9.ps}),
the inner blue component in  H$\alpha$ being
stronger than the red component, and the opposite being true for
 H$\beta$.
Comparing the line profiles and
 the widths of the inner  H$\alpha$ components
in the rms profile with those in the
difference profile 
(see Fig.~\ref{13463fg10.ps})
indicates that they are very similar, in contrast to the case for
the outer components.

The Balmer line profiles in Mrk\,926
 are extremely broad 
 (see Figs 8, 9). 
 We derived line-widths
(FWHM) of between 8~520 and 11~920~km~s$^{-1}$
from the mean and rms line profiles (Table\,6).
Furthermore, we parameterized the line widths of the rms profiles by
their line dispersions $\sigma_{line}$ (rms widths) (Fromerth \& Mellia
\cite{fromerth00}, Peterson et al. \cite{peterson04}).
The measured Balmer line dispersions $\sigma_{line}$
of more than 7~000~km~s$^{-1}$ (Table\,6)
are also indicative of very broad line widths.
The red wing of the H$\beta$ mean profile
 is heavily blended with the [\ion{O}{iii}] lines leading 
to a large uncertainty. 
The full widths at zero intensity (FWZI) in the Balmer lines
even correspond to 35~000 km s$^{-1}$.
\begin{table}
\centering
\caption{Balmer line widths: FWHM of the mean
and rms line profiles as well as line dispersion $\sigma_{line}$ (rms width)
of the rms profiles.}
\begin{tabular}{lccc}
\hline 
\noalign{\smallskip}
Line & FWHM (mean) & FWHM (rms) & $\sigma_{line}$ (rms) \\
     & [km s$^{-1}$] & [km s$^{-1}$] & [km s$^{-1}$] \\
(1)  & (2) & (3) & (4) \\ 
\noalign{\smallskip}
\hline
\noalign{\smallskip}
H$\beta$  & 11~920. $\pm$ 1000. & 8~620. $\pm$ 1000.  & 7~210. $\pm$ 350.\\
H$\alpha$ & 9~450. $\pm$ 200. & 8~520. $\pm$ 500. & 7~080. $\pm$ 400. \\
\noalign{\smallskip}
\hline 
\end{tabular}
\end{table}
%

\subsection{CCF and DCF analysis}
The size of the broad emission-line region in AGN can be estimated
by evaluating of the cross-correlation function (CCF) of the 
ionizing continuum flux light curve with the light curves
of the variable broad emission lines.
 An interpolation cross-correlation function
method (ICCF) was developed by Gaskell \& Peterson (\cite{gaskell87}). 
An independent method is
the discrete correlation function (DCF) method 
 described by Edelson \& Krolik (\cite{edelson88}).
Based on these papers, we developed our own ICCF and DCF code 
(Dietrich \& Kollatschny, \cite{dietrich95}).
 %
\begin{figure}
  \includegraphics[bb=40 60 570 780,width=63mm,angle=270] {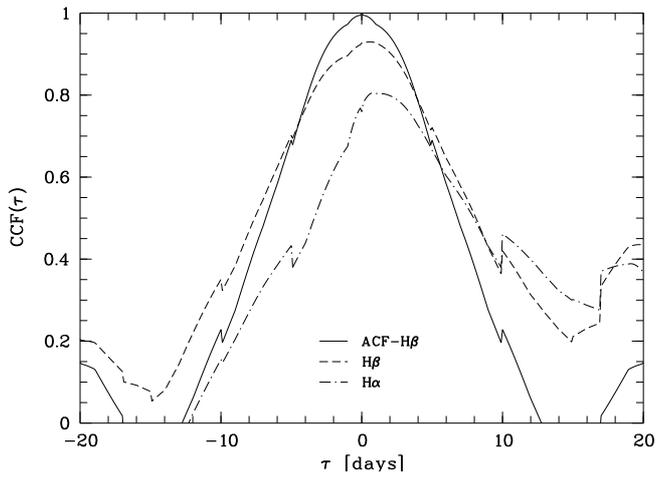}
\caption{Cross-correlation functions CCF($\tau$) of the continuum light
           curve at 5180\,\AA\ with the H$\beta$ and H$\alpha$ light curves
      as well as the H$\beta$ auto-correlation function.}
  \label{13463fg11.ps}
\end{figure}

We cross-correlated the individual continuum light curves of Mrk\,926
 with both the other continuum light curves and 
the light curves of the integrated H$\beta$ and H$\alpha$ lines 
using the ICCF method. Furthermore, we correlated the H$\beta$
light curve with itself (evaluating the auto-correlation function).
 Figures\,11, 12, and 13 show the cross-correlation functions CCF($\tau$)
of the continuum light curve at 5180\,\AA\ with the Balmer line
light curves and the H$\beta$ auto-correlation function (ACF).

We used 
the inner part within
$\pm$ 6000 km\,s$^{-1}$ only for the integrated H$\beta$ line
neglecting the outer line wings.
The outer blue line wing of H$\beta$ seems
to be contaminated with \ion{He}{ii}\,$\lambda$4686.
We derived the cross-correlation function 
of the H$\alpha$ line
for the same inner line profile segment to compare
these lines with each other.

The width of the H$\beta$  auto-correlation function provides an indication
of the BLR size (e.g. Gaskell \& Peterson (\cite{gaskell87}).
The H$\beta$ ACF shows that the BLR radius is smaller than 15--20 light-days
indicating that the BLR in Mrk\,926 is small.
Spatially resolved velocity delay maps of H$\beta$ and H$\alpha$ 
are discussed in Sect.~4.

We independently tested our ICCF results
by calculating the discrete correlation functions.
The results of the discrete correlation functions for  
H$\beta$ and H$\alpha$ (crosses with error bars)
 obtained with a binning of 5 days
are presented in Figs.~12 and 13.
For comparison, we present in Figs.~12 and 13
the cross-correlation functions CCF($\tau$) 
of the integrated  H$\beta$ and H$\alpha$ light
 curves (solid lines).
 The discrete correlation functions confirm the trend 
of a very small BLR. 

The centroids of the ICCF, $\tau_{cent}$,
were calculated using only the part of the CCF 
above 90\% of the peak value.
We derived the uncertainties in the cross-correlation results by
calculating the cross-correlation lags a large number of times using
a model-independent Monte Carlo method known as
``flux redistribution/random subset selection'' (FR/RSS),
a method described
by Peterson et al. (\cite{peterson98b}).
The error intervals correspond to 68\% confidence levels. 
%
%
\begin{figure}
 \includegraphics[bb=40 60 570 780,width=63mm,angle=270]{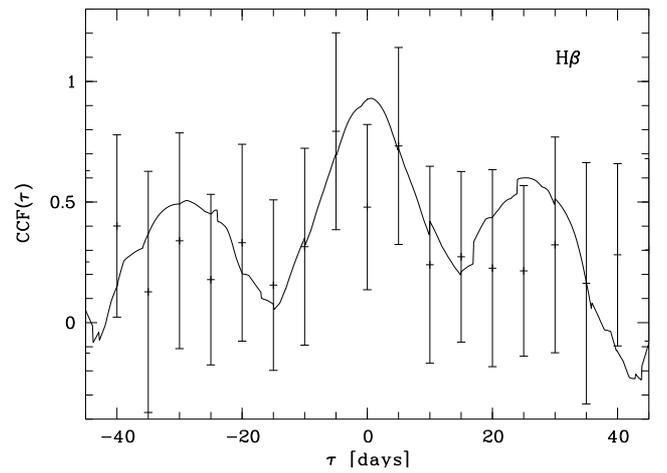}
  \caption{Cross-correlation function CCF($\tau$) (solid line)
  and DCF (error bars) of the H$\beta$ light
    curve with the 5180\,\AA\ continuum light curve.}
  \label{13463fg12.ps}
\end{figure}
%
%
%
\begin{figure}
 \includegraphics[bb=40 60 570 780,width=63mm,angle=270]{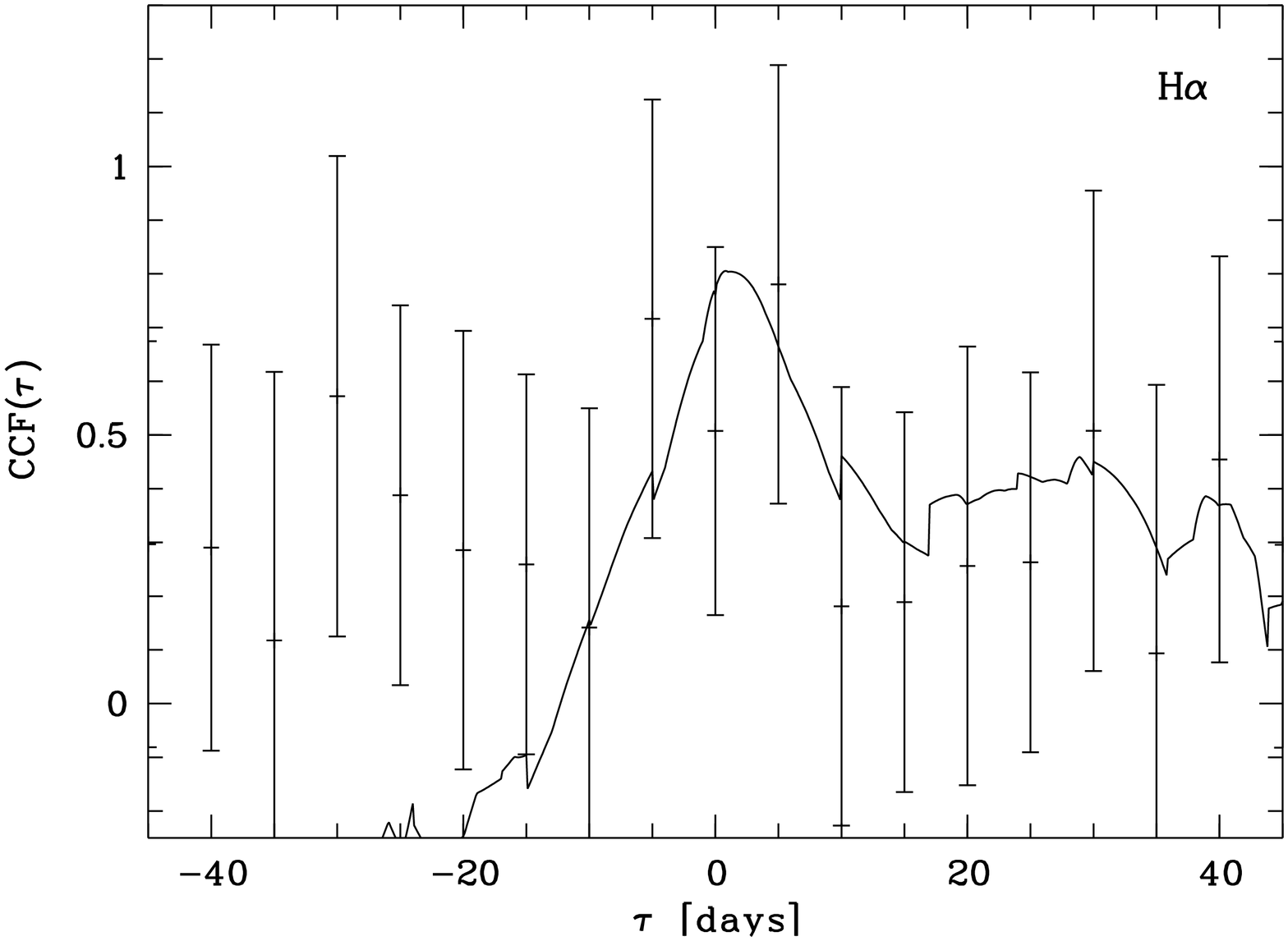}
  \caption{Cross-correlation function CCF($\tau$) (solid line)
  and DCF (error bars) of the H$\alpha$ light
    curve with the 5180\,\AA\ continuum light curve.}
  \label{13463fg13.ps}
\end{figure}
%
%

The final results of the ICCF analysis are given in Table\,7.
\begin{table}
\centering
\tabcolsep+16mm
\caption{Cross-correlation lags of the continuum light curve at 4600\,\AA\
            and of the Balmer-line light curves
     with respect to the 5180\,\AA\ continuum.}
\begin{tabular}{lc}
\hline 
\noalign{\smallskip}
Line & \multicolumn{1}{c}{$\tau_{cent}$} \\
     & \multicolumn{1}{c}{[days]}\\
(1)  & \multicolumn{1}{c}{(2)}\\
\noalign{\smallskip}
\hline
\noalign{\smallskip}
Cont.~4600          &   -$0.4^{+1.5}_{-1.5}$\\[.7ex]
H$\beta$            &   $0.5^{+1.5}_{-1.4}$\\[.7ex]
H$\alpha$           &   $1.6^{+1.9}_{-1.6}$\\[.7ex]
\noalign{\smallskip}
\hline 
\end{tabular}
\end{table}
The delay of
the integrated H$\beta$ line   
with respect to the continuum light
curve at 5180\,\AA\ 
corresponds to $0.5^{+1.5}_{-1.4}$~light-days only.
The delay of
the integrated H$\alpha$ line   
corresponds to $1.6^{+1.9}_{-1.6}$~light-days only.
The derived lags may be equal zero within the errors indicating that we have
derived at least an upper limit to the BLR size in Mrk~926.

The broad-line region size in Mrk~926 we derived from the Balmer lines
is small compared to those in other
Seyfert galaxies (for comparison see Kaspi et al. \cite{kaspi05}).
This is reviewed in more detail in the discussion section.

\subsection{Central black hole mass}

The central black hole mass in Mrk~926 can be derived from
the width of the broad emission line profiles based on the assumption 
that the gas dynamics are dominated by the central massive object,
by evaluating
$M = f\,c\,\tau_{cent}\,\Delta\,v^{2}\, G^{-1}  $.
%
The characteristic distance of the line-emitting region
 $\tau_{cent}$
 is given by the centroid of the individual cross-correlation
functions of the emission-line variations relative to the continuum variations
 (e.g. Koratkar \& Gaskell \cite{koratkar91}, Kollatschny \& Dietrich
\cite{kollatschny97}).
We derived upper limits to the H$\beta$ and H$\alpha$
cross-correlation lags $\tau_{cent}$  
based on the delay of the emission-line light curves with respect to the
continuum light curve at 5180\,\AA\
since a significant lag was not detected.\\       
The characteristic velocity $\Delta v$ of the emission-line
region can be estimated from either the FWHM of the rms profile or from
the line dispersion $\sigma_{line}$.
We wish to compare the central black hole mass in Mrk~926 with the
central black hole masses of other AGN
 in the database of Peterson et al. (\cite{peterson04}).
Therefore,
we used the line dispersion $\sigma_{line}$ to
calculate the central black hole mass in Mrk~926.

The scaling factor $f$ in the equation above is of the order
of unity and depends
on the kinematics, structure, and orientation of the BLR.
The scaling factor may differ from galaxy to galaxy e.g. if we see the
central accretion disk including the BLR from the edge or face-on. 
Again we recall that we wish to
compare the central black hole mass in Mrk~926 with the
central masses in the AGN sample 
 of Peterson et al. (\cite{peterson04}), and therefore,
we adopt their mean value of f=5.5.\\
In Table\,8, we list our virial mass calculations of the central black hole
in Mrk~926 based on the Balmer lines.
\begin{table}
\centering
\tabcolsep+4.5mm
\caption{
Line dispersion $\sigma_{line}$ (rms width)
of the rms Balmer profiles,
upper limits of the H$\beta$ and H$\alpha$ emission line cross-correlations,
 and derived central black hole mass limits.
}
\begin{tabular}{lcccc}
\hline 
\noalign{\smallskip}
Line & $\sigma_{line}$ (rms) & $\tau_{cent}$ & $M$ \\
     & [km s$^{-1}$] & [days] & [$10^7 M_{\odot}$]\\
(1) & (2) & (3) & (4)\\ 
\noalign{\smallskip}
\hline
\noalign{\smallskip}
H$\beta$  & 7~210. $\pm$ 350. & $\leq$2.0  & $\leq$11.2 $\pm$ 0.001 \\
H$\alpha$ & 7~080. $\pm$ 400. & $\leq$3.5  & $\leq$18.8 $\pm$ 0.002 \\
\noalign{\smallskip}
\hline 
\end{tabular}
\end{table}
Based on H$\beta$, we derive an upper limit to the black hole mass of 
\[ M\leq 11.2 \times 10^{7} M_{\odot} . \]
Based on H$\beta$ and H$\alpha$, we derive a weighted average limit of
\[ M\leq 15.0 \pm 5.4 \times 10^{7} M_{\odot}  \]
for Mrk\,926.

\section {2-D CCF of the H$\alpha$ and H$\beta$
       line profiles}

We now investigate in detail the line profile variations of H$\alpha$
and H$\beta$.
We proceed in the same way as when we studied the line profile variations in
Mrk\,110 (Kollatschny \& Bischoff  \cite{kollatschny02}, Kollatschny
\cite{kollatschny03}).  

We sliced the H$\alpha$ and H$\beta$ velocity profiles  
into velocity segments of widths $\Delta$v = 400 km s$^{-1}$
for all the spectra taken in 2005. The value of
400 km s$^{-1}$ corresponds to the spectral resolution of our observations.
We measured the intensities of all subsequent
velocity segments 
from v = -15~000 until +15~000 km s$^{-1}$ and compiled their light curves.
The intensity of
the central line segment was integrated from
v = -200 until +200 km s$^{-1}$.

All the light curves of the line segments including the light curves of
the continuum  at 5180\,\AA{} look similar.
There are no clear-cut differences between the outer and inner segments
or between the red and blue wings. 
We computed cross-correlation functions (CCFs) of all
line segment ($\Delta$v = 400 km s$^{-1}$) light curves
with the 5180\,\AA{} continuum light curve.

The derived delays of the segments with respect to the 5180\,\AA{}
continuum light curve are shown in 
Figs.~14 and 15 as a function of distance to the line center.
%
%
%
\begin{figure}
 \hbox{
\includegraphics[bb=36 149 380 643,width=56mm]{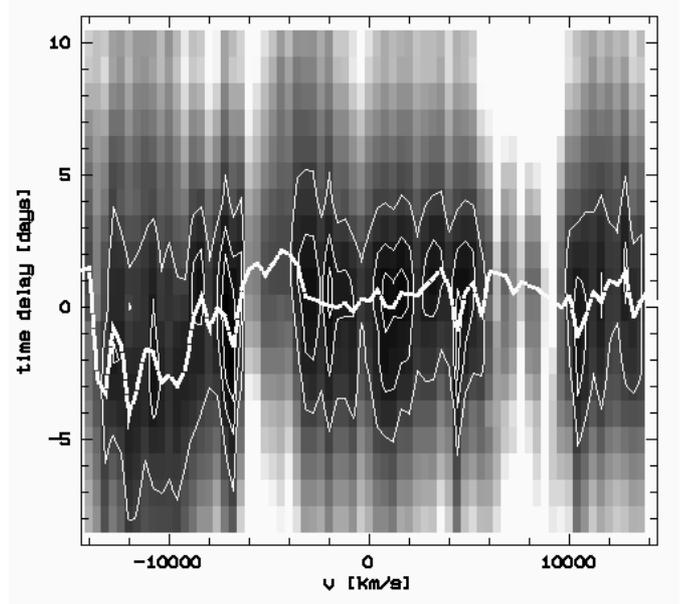}
}
  \caption{The 2-D CCF($\tau$,$v$) shows the correlation of the H$\beta$ line  
segment light curves with the continuum light curve at  5180\,\AA{}
as a function of velocity and time delay (grey scale).
Contours of the correlation coefficient are overplotted at levels of 0.93,
0.89, and 0.80 (solid lines). The heavy dashed line connects the centers of all
individual cross-correlation functions.
}
   \label{13463fg14.ps}
\end{figure}
%
%
\begin{figure}
\hbox{
\includegraphics[bb=36 149 380 643,width=56mm]{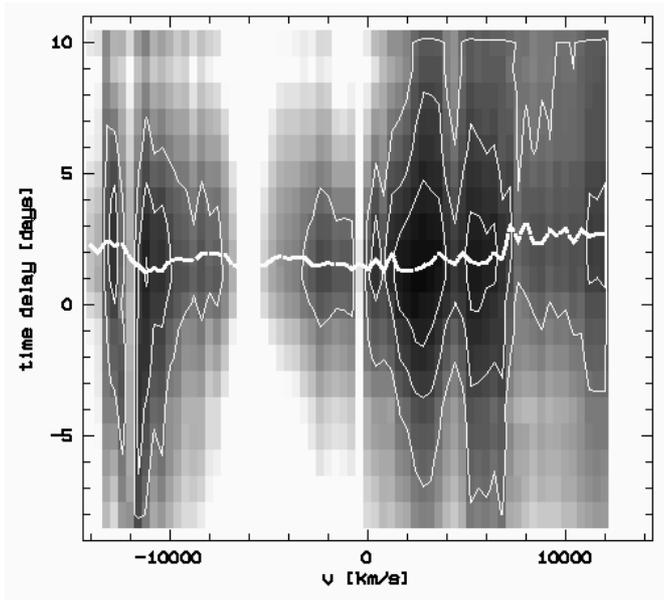}
}
  \caption{The 2-D CCF($\tau$,$v$) shows the correlation of the H$\alpha$ line  
segment light curves with the continuum light curve at  5180\,\AA{}
as a function of velocity and time delay (grey scale).
Contours of the correlation coefficient are overplotted at levels of 0.85, 
0.75, and 0.65 (solid lines).The heavy dashed line connects the centers of all
individual cross-correlation functions. 
}
   \label{13463fg15.ps}
\end{figure}
These 2-D CCFs are presented in gray scale
for the H$\beta$ and H$\alpha$ lines.
We calculated the CCFs of the H$\alpha$ line segments
only for segments blue-wards of v =12~400 km s$^{-1}$ because
the atmospheric lines are
contaminate the outer red wing of H$\alpha$ significantly.

The thin solid lines of Figs.~14 and 15 delineate the
contour lines of the correlation coefficient
at levels of .93, .89, and .75 for H$\beta$ and at levels of .85, .75,
and .65 for H$\alpha$.
The heavy dashed line connects the centers of all
cross-correlation functions for the different line profile segments. 
The time delay of the line segment light curves was calculated from the
uppermost 10 percent of the cross-correlation functions.
Only  correlation coefficients with values greater than .6 
were considered for H$\beta$.

Figure~16 shows the delay of the individual line segments together for
both lines H$\alpha$ and H$\beta$; these delays correspond to the heavy dashed
lines in Figs.~14 and 15.
%
%
\begin{figure}
\includegraphics[width=63mm,angle=270]{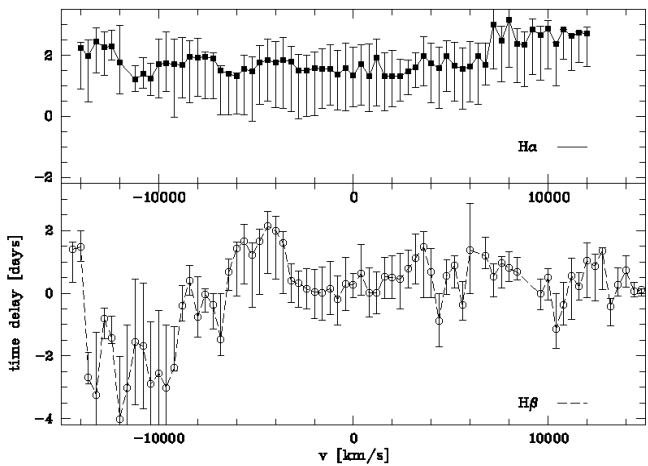}
  \caption{Time delay of the individual line segments of 
    H$\alpha$ (solid line, filled square) and  H$\beta$ (dashed line, open
 circle) with respect to the continuum light curve at  5180\,\AA{}.}
  \label{13463fg16.ps}
\end{figure}
We now consider the relative delays of the individual
line segments for placing constraints on the velocity and/or geometry in the
BLR of Mrk\,926.
We therefore indicate in Fig.~16 errors based on the flux
redistribution method only. In this case, we did not consider additional
random subsets of the light curves (Peterson et al. \cite{peterson98b}).
This would have caused an additional error of about $\pm{}$ 1 light-day for the
line segments.\\ 

Both lines (H$\alpha$ and H$\beta$)
respond with very short delays relative to the continuum
at 5180\,\AA{} as derived previously from the integrated
lines. Within the small-scale variations, they are nearly constant
along the line profile with one exception: There is a drop in the delay
of the outer blue wing of H$\beta$ between v =-14~000 and -8~400 km s$^{-1}$. 
This wavelength range 
 corresponds exactly to the \ion{He}{ii}\,$\lambda$4686 line
centered on -10~800 km s$^{-1}$.  
It is known from other galaxies (e.g. Mrk\,110, Kollatschny et al.
\cite{kollatschny01}) that the \ion{He}{ii}\,$\lambda$4686 line exhibits
shorter delays with respect to the ionizing continuum than the
 H$\beta$ line. The apparent negative delay of two light-days with respect
to the continuum light curve
at 5180\,\AA{} might be caused by a general
 systematic error or by an alternative
possibility
that the  continuum flux at 5180\,\AA{} does not correspond to the
ionizing continuum flux.\\

Figure~17 shows the maximum response of the H$\alpha$ and H$\beta$
line segment cross-correlation functions. 
%
%
\begin{figure}
\vbox{
 \includegraphics[bb=40 60 570 780,width=63mm,angle=270]{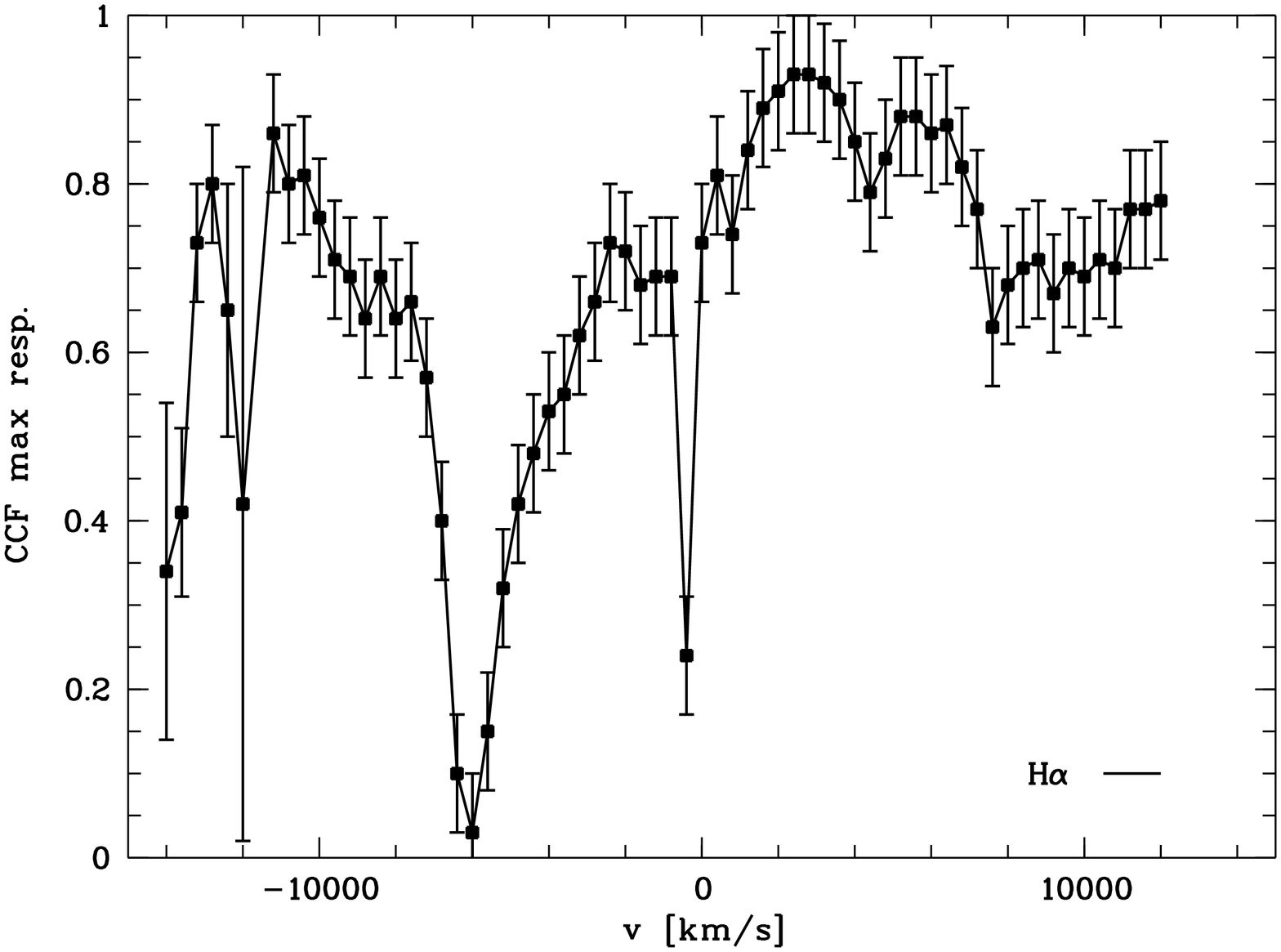}\\
 \includegraphics[bb=40 60 570 780,width=63mm,angle=270]{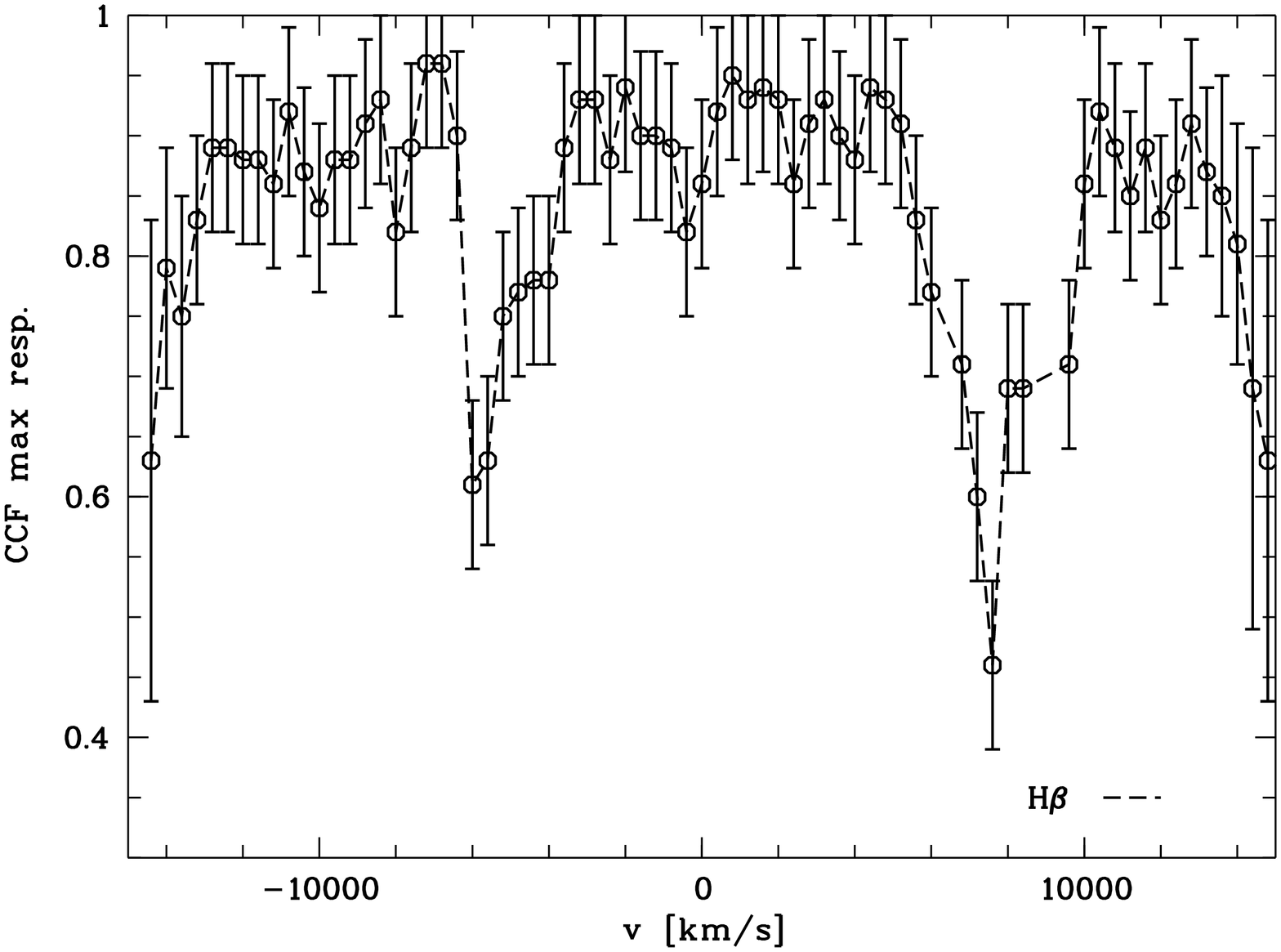}
}
  \caption{Maximum response of 
the correlation functions of the
 H$\alpha$ (solid line, filled square) and  H$\beta$ (dashed line, open
 circle)
line segment light curves with the continuum light curve at  5180\,\AA{}
.}
  \label{13463fg17.ps}
\end{figure}
It is surprising that the characteristics of the
H$\alpha$ and  H$\beta$ line segment responses
have the same pattern.
The response is independent
of the intensity of the line segments.
 There are two pronounced outer minima at 
 -6~000 and +7~600 km s$^{-1}$ in the response distribution as well as 
central minima at -400 and +800 km s$^{-1}$.
The intervals of the five distinct components in
the response distribution of H$\beta$ and H$\alpha$ are given in Table~9.
\begin{table}
\tabcolsep+8.5mm
\centering
\caption{Intervals of the response segments derived from the 
 2-D CCFs of H$\beta$ and H$\alpha$ .} 
\begin{tabular}{l D{t}{\,\textrm{to}\,}{7,7}}
\hline 
\noalign{\smallskip}
Interval & \multicolumn{1}{c}{Response sections of H$\alpha$, H$\beta$}\\
        & \multicolumn{1}{c}{[km s$^{-1}$]} \\
 \multicolumn{1}{c}{(1)}  &  \multicolumn{1}{c}{(2)} \\ 
\noalign{\smallskip}
\hline
\noalign{\smallskip}
outer blue & -14~000 t - 6~000 \\
inner blue &  -6~000 t - 400 \\
central    &   -400 t  ~~800  \\ 
inner red  &   800 t ~~7~600 \\ 
outer red  &   ~7~600 t ~~14~800 \\
\noalign{\smallskip}
\hline 
\end{tabular}
\end{table}
The additional
drop in the H$\alpha$ response close to 
-12~000 km s$^{-1}$ is caused by the very broad
cross-correlation function of this line segment
that has no clear maximum (see Fig.~14).
This response value is therefore
affected by a large error.

\section{Discussion}

Mrk\,926 is a very broad-line Seyfert 1 galaxy.
It remains unclear the extent to which
 the extreme emission line-widths
in the individual very broad-line AGN are caused
by intrinsic properties of the galaxy or
by projection effects of the central BLR accretion disk.

\subsection{Mean and rms line-profiles in Mrk~926}

The mean and the rms Balmer lines in Mrk\,926 
are extremely broad. They have full width half maxima (FWHM)
of the order of 10~000 km s$^{-1}$
(Figs.~\ref{13463fg7.ps} and
\ref{13463fg8.ps}).
The full width at zero intensity (FWZI)
of  H$\beta$ amounts to 35000 km s$^{-1}$.

It was noted by Fromerth \& Mellia
(\cite{fromerth00}) and Peterson et al. (\cite{peterson04}) that an
emission line width can be parameterized by their FWHM and their
line dispersion $\sigma_{line}$.
We derived a line dispersion $\sigma_{line}$ of 7~100 km s$^{-1}$
for the Balmer lines in Mrk\,926.
No other galaxy
in the sample of 35 AGN (Peterson et al. \cite{peterson04})
exhibits such a broad line dispersion $\sigma_{line}$.
In terms of the FWHM, only 3C~390.3 shows broader wings than Mrk\,926. 

For a Gaussian line profile, the ratio FWHM/$\sigma_{line}$ is 2.355.
Peterson et al. (\cite{peterson04}) derived a mean ratio
of FWHM/$\sigma_{line}$=2.03
in their sample indicating that 
their emission lines
had in general weaker cores and stronger wings than Gaussian profiles. 
We derived an extreme  FWHM/$\sigma_{line}$ ratio
of 1.2 only for Mrk\,926
based on the rms Balmer profiles. This ratio confirms the
existence of the very strong line wings seen in Fig.~9.
Only five AGNs in the sample of
Peterson et al. (\cite{peterson04}) have similar ratios of
FWHM/$\sigma_{line}\leq$1.2 for H$\beta$  (their Table 6, Fig.~9).

The rms profiles in Mrk\,926 are more complex than
one would expect if they were formed in a basic optically thin accretion disk.  
A simple accretion disk would produce a double-peaked profile
(e.g.  Welsh \& Horne \cite{welsh91}, Horne et al. \cite{horne04}),
although this simple model cannot easily explain
 the existence of two inner
and two outer line components in addition to a central component.
Furthermore, the components are not arranged symmetrically with respect to
v = 0  km s$^{-1}$.
In a few other broad-line galaxies (3C390.3, 3C332), similar
complex structures have been 
found in the rms profiles (Gezari et al. \cite{gezari07}).
The geometrical and physical conditions causing
these complex profiles are not yet fully understood.

The different behaviour of the line components
in  H$\alpha$ and H$\beta$ 
(Figs.~\ref{13463fg6.ps} to \ref{13463fg9.ps})
indicates that the components might originate in different regions or in
regions of different optical thickness.
The segmented structure in the line profiles
will be addressed again in conjunction with 
the two-dimensional cross-correlation function (Sect. 5.4).

\subsection{The BLR size in Mrk~926}

We correlated the individual continuum light curves of Mrk\,926
 with each other and with the Balmer emission-line light curves.
Figures~11 to 13 and Table~7 display the
delays of the Balmer lines
with respect to the continuum light curve at 5180\,\AA{}
and the H$\beta$ auto-correlation function.

Mrk\,926 has a very small BLR. This is based on the
H$\beta$ auto-correlation function
as well as on the delays
of the Balmer line light
curves with respect to the continuum light curve at 5100\,\AA{} .  
We determined a line-averaged H$\beta$ BLR size of
$0.5^{+1.5}_{-1.4}$~light-days
and a H$\alpha$ BLR size of
$1.6^{+1.9}_{-1.6}$~light-days.
The upper limit of 2~light-days to the H$\beta$
 BLR size in Mrk\,926 corresponds to the smallest BLR sizes
in the AGN sample of Kaspi et al. (\cite{kaspi05}).
They investigated the relationship between
AGN luminosity and broad-line size
on the basis of the most accurate
determinations of BLR radii for 35 AGNs
(their Fig~2).
Mrk\,926 deviates slightly from the general relationship between luminosity
and broad-line region size in AGNs.
However, other broad-line Seyfert galaxies of comparable luminosity
 e.g. IC\,4329A, NGC\,7469, and NGC\,4593
also exhibit similar deviations from the general trend.

From the general relation
of Kaspi et al. (\cite{kaspi05}),
one would predict a larger BLR size of Mrk\,926
with respect to the measured continuum luminosity. 
A continuum luminosity of
$\lambda{}\,$L$_{\lambda}(5100\,\AA{}) = 5.91 \pm{} 0.56 \times 10^{43}\,$erg\,s$^{-1}$
corresponds to a seven times larger BLR size of
R(BLR)= 15.51 light-days
using Eq. (6) of 
Kaspi et al. (\cite{kaspi05}). 
Such a large BLR size can be excluded by the auto-correlation and
cross-correlation functions.

There is the possibility that the derived small delay
for the Balmer-line emitting region does not correspond to the
true distance of the line-emitting region from the central ionizing source.
This distance might be larger by the order of two to four light-days
because of the possibility that the observed continuum at 5180\,\AA{} does
not correspond
to the ionizing continuum in the UV below 1000\,\AA{}. 
Evidence of a wavelength-dependent continuum delay in the optical
was found 
in the Seyfert galaxy NGC~7469 (Wanders et al. \cite{wanders97},
Collier et al. \cite{collier98}). 
These observations were interpreted as reprocessing of high-energy radiation 
close to the center
in an irradiated black-body accretion disk with a 
  $T \propto R^{-4/3}$ temperature profile, which causes delays of
 $\tau \propto \lambda^{4/3}$.
Sergeev et al. (\cite{sergeev05})
published wavelength-dependent continuum delays based on broad-band
photometry of the light curves of 14 AGN. They interpreted their
observations in terms of reprocessed emission of an X-ray source above the
accretion disk.
Gaskell (\cite{gaskell07}, \cite{gaskell08}) explained the observed
wavelength-dependent continuum delays as possibly being caused by
contamination of an intrinsically coherently variable continuum
with the Wien tail of
the thermal emission from the hot dust in the surrounding torus. The
detection of a lag in the polarized flux provides a way to discriminate
between both models.


\subsection{Central black hole mass in Mrk~926}

We derived an upper limit of
$M= 11.2 \times 10^{7} M_{\odot}$
to the central black hole mass
based on both the distance of the H$\beta$ line-emitting region and the
line dispersion $\sigma_{line}$ of the rms line profile.
Peterson et al. (\cite{peterson04}) presented a black hole mass-luminosity
relationship for 35 reverberation mapped AGNs.
With respect to its optical luminosity, the black hole mass of Mrk~926
 is within the
expected range of a few times $10^{7} M_{\odot}$ (their Fig.\,16).  

An estimate of the black hole mass based on the line width
alone is less reliable.
From the line widths of  Mrk~926,
 O'Neill et al. (\cite{oneill05}) derived a very high
black hole mass of $3.5 \times 10^{8} M_{\odot}$
and compared this number with the observed X-ray variability amplitude
observed by ASCA.
But they derived only an upper limit to their excess
variance (see their Fig.\,2). Since the intrinsic black hole mass in Mrk\,926
is lower by a factor of three, one would expect a slightly more
variable X-ray source.  


\subsection{2D velocity map}

The delays of the individual line segments of H$\alpha$ and H$\beta$
 with respect to the continuum variability at 5180\,\AA{}
are very short
(0 to 3 light-days only) and more or less constant over the entire
line profile (Figs.~14 to 16).  
There is a drop in 
the outer blue wing of H$\beta$ 
 between $v$=-14~000 and -8~400 km s$^{-1}$. 
This line region corresponds to a blend of the
 \ion{He}{ii}\,$\lambda$4686 line in the  blue wing of H$\beta$.
The \ion{He}{ii}\,$\lambda$4686 line
is centered on -10~800 km s$^{-1}$ with respect to H$\beta$.  
The H$\alpha$ line does not show this drop in the blue wing.

The computed delay of the \ion{He}{ii}\,$\lambda$4686 line is 
negative 
with respect to the continuum at 5180\,\AA{} by about two light days. 
This negative delay is within our
systematic error of three light-days. Another explanation of the negative delay
is the possibility that
the optical continuum light curve at 5180\,\AA{} does not correspond to 
the ionizing continuum of the \ion{He}{ii}\,$\lambda$4686 line in the far-UV
because of a wavelength-dependent continuum delay. Spectral observations
on a daily basis are needed to solve this question
that the \ion{He}{ii}\,$\lambda$4686 line
originates in the
innermost region while the H$\alpha$ line originates 
in more exterior regions than H$\beta$
as has been found e.g. in Mrk~110
(Kollatschny et al. \cite{kollatschny01}) too. 

Figures~\ref{13463fg14.ps} and \ref{13463fg15.ps} show
the 2-D cross-correlation functions CCF($\tau$,$v$) of
 the H$\beta$ and H$\alpha$ line  
segment light curves with the continuum light curve at  5180\,\AA{}
as a function of velocity and time delay (grey scale).
These 2-D CCF($\tau$,$v$) are
mathematically very similar to a 2-D response function $\Psi$ (Welsh
 \cite{welsh01}).
We
compare our observed velocity-delay pattern with model calculations of
 Welsh \& Horne (\cite{welsh91}), 
Perez et al. (\cite{perez92}),  O'Brien et al. (\cite{obrien94}),
Horne et al. (\cite{horne04}), 
and with results of Mrk\,110 (Kollatschny, \cite{kollatschny03}).


The delay in the individual line segments 
is very small. We could not discern any trend as a function of distance to the
line center except for the drop in the blue H$\beta$ wing.
The observed uniformity
 of the line-segment delays in Mrk\,926
 is different from that observed in Mrk\,110
(Kollatschny, \cite{kollatschny03}).
In that galaxy, the emission line wings responded far more quickly 
than the central line region.
Their central Balmer line
regions responded with a delay of about 30 light days
with respect to the ionizing continuum.
However, an existing pattern
might be hidden in the 2-D CCFs of Mrk\,926
 due to the very small BLR. Our temporal sampling is insufficient
to detect any detailed pattern on timescales shorter than about three
light-days.

The observed emission lines and their corresponding 2-D CCFs are very broad.
The line widths of Mrk\,926 amount to 10~000 km s$^{-1}$ (FWHM) compared to
line widths of 2~000 km s$^{-1}$ only in Mrk\,110
(Kollatschny et al. \cite{kollatschny01}). It has been demonstrated that
the narrow Balmer and helium
 emission lines in Mrk\,110 originate in an accretion disk seen nearly
face on  (Kollatschny \cite{kollatschny02}, \cite{kollatschny03}).
The response of the very broad emission lines in Mrk~926 might 
be caused by an accretion disk seen nearly edge on 
relative to e.g. the model calculations of Perez et al.
(\cite{perez92}, their Fig.~6)
or Welsh \& Horne (\cite{welsh91}, their Fig.~5).

Figure~18 shows the response of the H$\alpha$ and H$\beta$
line segment cross-correlation functions as well as their normalized
rms profiles in one plot. The response curves
have been shifted by 0.5 to avoid any strong overlap of the curves. 
Furthermore, two spikes in the [OIII] residuals have been removed.
\begin{figure}
\vbox{
 \includegraphics[bb=40 60 570 780,width=63mm,angle=270]{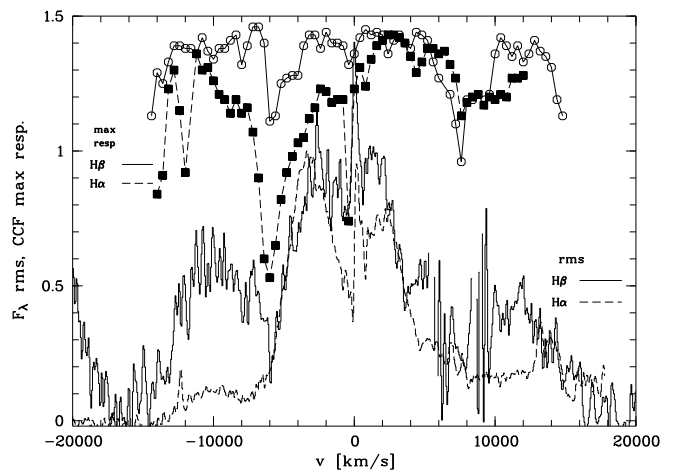}
}
\caption{Maximum response of 
the correlation functions of the
 H$\alpha$ (solid line, filled square) and  H$\beta$ (dashed line, open
 circle)
line segment light curves with the continuum light curve at  5180\,\AA{}
as well as their normalized rms profiles. The response curves
are shifted by 0.5 to avoid strong overlap of the curves.}
  \label{13463fg18.ps}
\end{figure}
It is interesting that the double structure in the  H$\alpha$ and H$\beta$
response curves -- 
which contains two separate inner and outer components (Table 9) --
is also evident in the rms profiles (Table 5), which were derived in
a completely different manner.

The response of the (inner) red wing is stronger than the response of the
(inner) blue wing  (Figs.~\ref{13463fg14.ps}, \ref{13463fg15.ps}, 
and \ref{13463fg17.ps}).
The stronger response of the red line wing relative to the 
blue line wing was predicted by Chiang \& Murray (\cite{chiang96})
in their accretion disk-wind model of the BLR.
Similar evidence of an accretion disk wind
was detected in Mrk\,110 ( Kollatschny \cite{kollatschny03}).
The response of the outer line wings (shortwards of $-6~000$ km s$^{-1}$
and longwards of
$8~000$ km s$^{-1}$)  does not follow this trend. The
outer line wings might originate from
different physical conditions.

A structured  response may be considered as independent evidence of
a structured BLR:
the broad emission lines may then
 originate in a homogeneous medium or
different line segments
may originate in different regions and/or under different
physical conditions.

\section{Summary}

A spectroscopic monitoring campaign of the very broad line
AGN Mrk\,926 was carried out by ourselves with the 9.2m Hobby-Eberly Telescope
in the years 2004 and 2005.
The main results of our study can be summarized as follows:\\

1. The rms profiles of the very broad Balmer lines (10~000 km s$^{-1}$ FWHM)
are structured showing two inner
and two outer line components in addition to a central component.
The outer and inner line segments vary with different amplitudes.\\

2. The radius of the BLR is quite small 
with an upper limit of 2~light-days for the H$\beta$
BLR size.\\

3. We calculated an upper limit of
$M= 11.2 \times 10^{7} M_{\odot}$
for the central black hole mass in Mrk\,926.
This result was based on the  H$\beta$ line dispersion $\sigma_{line}$ and
the upper limit to the
cross-correlation lag $\tau_{cent}$. 
The central black hole mass in Mrk~926
 is inside the
expected range of a few times $10^{7} M_{\odot}$ 
with respect to its optical luminosity.\\

4. The 2-D cross-correlation functions CCF($\tau$,$v$) of H$\beta$ and
H$\alpha$ are flat within the error limits.
However, the response of the Balmer line
segments is structured.
A double structure in the  H$\alpha$ and H$\beta$
response curves -  showing two separate inner and outer components -
has also been seen in the rms line profiles. The different line segments
might originate in separate regions and/or under different physical conditions.

\begin{acknowledgements}
We thank Shai Kaspi and Martin Gaskell for useful comments.
This work has been supported by the Niedersachsen - Israel Research
Cooperation Program ZN2318.

\end{acknowledgements}

\newpage
\end{document}